\documentclass{aa}

\usepackage[varg]{txfonts}
\usepackage{natbib}
\usepackage{rotating}
\usepackage{amsmath}
\usepackage{graphicx}
\usepackage{textcomp}
\usepackage{xcolor}
\usepackage{tabularx}
\usepackage{ulem} 

\begin{document}

\title{Vortex-like kinematic signal, spirals, and beam smearing effect in the HD~142527 disk}

\author {Y. Boehler \inst{1} \and  
         F. M\'enard \inst{1} \and   
         C.M.T. Robert \inst{1} \and 
         A. Isella \inst{3} \and
         C. Pinte \inst{4} \and
         J.-F. Gonzalez \inst{5} \and
         G. van der Plas \inst{1} \and 
         E. Weaver \inst{3} \and
         R. Teague \inst{6} \and
         H. Garg \inst{4} \and 
         H. M\'eheut \inst{2} }
         
\institute{Univ. Grenoble Alpes, CNRS, IPAG, F-38000 Grenoble, France \\
       \and Universit\'e Côte d'Azur, Observatoire de la Côte d'Azur, CNRS, Laboratoire Lagrange, Bd de l'Observatoire, CS 34229, 06304 Nice cedex 4, France \\
       \and Rice University, Department of Physics and Astronomy, Main Street, 77005 Houston, USA \\
       \and Monash University, Wellington Rd, Clayton VIC 3800, Australie \\
       \and Univ Lyon, Univ Claude Bernard Lyon 1, Ens de Lyon, CNRS, Centre de Recherche Astrophysique de Lyon UMR5574, F-69230, Saint-Genis-Laval, France \\
       \and Center for Astrophysics, Harvard \& Smithsonian, 60 Garden Street, Cambridge, MA 02138, USA}      
       
\abstract
{Vortices are one of the most promising mechanisms to locally concentrate millimeter dust grains and allow the formation of planetesimals 
through gravitational collapse. The outer disk around the binary system HD~142527 is known for its large horseshoe structure with 
azimuthal contrasts of $\sim$ 3-5 in the gas surface density and of $\sim$ 50 in the dust. Using $^{13}$CO and C$^{18}$O J = 3-2 
transition lines, we detect kinematic deviations to the Keplerian rotation, which are consistent with the presence of a large vortex 
around the dust crescent, as well as a few spirals in the outer regions of the disk. Comparisons with a vortex model suggest 
velocity deviations up to 350 m s$^{-1}$ after deprojection compared to the background Keplerian rotation, as well as an extension 
of $\pm$ 40 au radially and $\sim$ 200$^\circ$ azimuthally, yielding an azimuthal-to-radial aspect ratio of $\sim$ 5. Another 
alternative for explaining the vortex-like signal implies artificial velocity deviations generated by beam smearing in association with 
variations of the gas velocity due to gas pressure gradients at the inner and outer edges of the circumbinary disk. The two scenarios are 
currently difficult to differentiate and, for this purpose, would probably require the use of multiple lines at a higher spatial resolution. 
The beam smearing effect, due to the finite spatial resolution of the observations and gradients in the line emission, should be common in 
observations of protoplanetary disks and may lead to misinterpretations of the gas velocity, in particular around ring-like structures.}


\maketitle

\section{Introduction}
\label{sec:Intro}

For a couple of years, observations with the Atacama Large Millimeter Array (ALMA) have detected numerous deviations to the Keplerian 
rotation in protoplanetary disks, most of which are consistent with the presence of embedded Jupiter-mass planets interacting with the 
disk \citep{Teag2018a, Pint2018, Casa2019b}. The first detections were found in the multi-ringed disk around HD~163296. Using 
azimuthal averaging, \cite{Teag2018a} studied the gas kinematics which is sensitive to radial pressure gradients and therefore a 
direct probe for local variations of the gas surface density. With this method, they found gaps in the gas, possibly carved by 
embedded planets and colocated with gaps previously discovered in the dust. \cite{Pint2018} found a local deviation, also called 
velocity kink, in the outer regions of the gaseous disk at $\sim$ 260 au, which they interpreted as a perturbation due to a 
massive planet. Since then, numerous kinks, wiggles, and Doppler flips have been discovered in other protoplanetary disks 
\citep{Pint2019, Casa2019b, Pint2020}, suggesting the presence of numerous planets.

Regarding their formation, theories and simulations have long shown that large dust grains (i.e., $\geq$ 10-100 microns) 
must be trapped in local pressure maxima to avoid processes such as inward radial drift, bouncing, and grains fragmentation 
\citep{Weid1977, Birn2010}. When the dust-to-gas ratio in these traps reaches a value between 0.1 and 1 (with an initial value of 0.01), 
hydrodynamic instabilities such as the streaming instability can be triggered and potentially lead to the formation of planetesimals 
\citep{Youd2005, Joha2007, Bai2010, Raet2015, Auff2018}.

The process of dust trapping is likely common in protoplanetary disks. Comparisons of dust ring structures with gas emission suggest that 
dust grains are trapped in radial gas pressure maxima in several disks \citep{Dull2018, Roso2020}. Evidence of both 
radial and azimuthal dust trapping has also been found in approximately ten protoplanetary disks so far. These dust concentrations can 
present large variations in magnitude, be single or multiple. They can be located in dust rings such as in V1247~Orionis \citep{Krau2017}, 
MWC~758 \citep{Boeh2018, Dong2018, Casa2019a}, HD~143006 \citep{Andr2018}, and HD~135344B \citep{vdM2016, Cazz2018}, or they can be 
characterized by a large horseshoe structure such as in HD~142527 \citep{Casa2013, Muto2015, Boeh2017, Soon2019}, IRS~48 \citep{vdM2013, 
Calc2019}, AB Aur \citep{Tang2012}, and SZ91 \citep{vdM2018}. 

Observations of these asymmetries at multiple wavelengths indicate that grains with a larger size a, proportional to $\lambda$/(2$\pi$), 
have a more compact spatial distribution \citep{vdM2015, Casa2019a}. Indeed, the efficiency of the dust trapping depends on the Stokes number 
(i.e., the ratio of the dust stopping time to their orbital period), which is proportional to the grain size over gas density. The 
strongest effect of drag is expected for a Stokes number of $\sim$ 1, while dust grains probed with ALMA have a Stokes number generally 
estimated around $\sim$ 10$^{-2}$-10$^{-1}$. It is therefore likely that the dust trapping process is much more frequent than currently 
observed with ALMA. Its study will require longer wavelengths to probe grains with a Stokes number close to 1, even in dense 
gas areas \citep{vdM2020}, and to be less limited by the dust optical depth.

One of the main theories regarding the production of azimuthal dust concentrations invokes anticyclonic vortices \citep{Barg1995, Lyra2013, 
Baru2016, Sier2017}, which can form via the Rossby wave instability \citep{Love1999, Li2000}, for instance in a steep gradient in density 
\citep{deVa2007, Zhu2014} or in viscosity \citep{Varn2006, Rega2012}. The edge of a disk cavity or of a ring is thus a 
favorable site for vortex formation. However, these vortices also require a low viscosity to form, with a turbulent parameter $\alpha 
\lesssim$ 10$^{-3}$ \citep{Shak1973, Zhu2014}. 

More recently, numeric simulations have also shown that a companion with a mass ratio $q$ $\geq$ 0.05 can create an eccentric cavity 
and trigger an azimuthal clump in the gas \citep{Shi2012, Ragu2017}. For even larger mass ratios, the azimuthal gas contrast may reach 
a value between two and four in steady state, similar to what is observed in HD~142527 \citep{Pric2018, Ragu2020}. 
Performed with a relatively high turbulent viscosity ($\alpha$ $\sim$ 5 $\times$ 10$^{-3}$), these simulations did not produce 
vortices but were still able to trap dust particles in the gas clump, which is rotating at Keplerian velocity.

Apart from the presence of a massive companion, the main physical property that may favor one or the other scenario is thus the gas 
viscosity. With the level of turbulence in disks being poorly known \citep{Flah2020}, it is unclear whether vortices can actually 
develop. Only the detection of their kinematic signature will undoubtedly confirm their presence. The study of the gas kinematics 
is therefore the key element for understanding the origin of the dust concentrations in protoplanetary disks.

  The binary system HD~142527 is located at 157 $\pm$ 2 pc, based on the stellar parallax \citep{Gaia2018}, and consists of a $\sim$ 
2.1 $\mathrm{M}_\sun$ Herbig star and of a $\sim$ 0.3 $\mathrm{M}_\sun$ companion in an eccentric and non-coplanar orbit, currently at 
a distance of about 12 au from the central star \citep{Bill2012, Clau2019}. It is surrounded by a bright circumbinary disk with an extremely large 
extent, from 100 au to 300 au in the dust emission and up to 800 au in the gas emission. Observations and models of the outer disk 
reveal a horseshoe morphology with an azimuthal contrast of $\sim$ 3-5 in the gas surface density and of $\sim$ 50 in the dust surface 
density \citep{Muto2015, Boeh2017}, leading to a gas-to-dust ratio of $\sim$ 1 at the north of the disk \citep{Yen2020}. 

Moreover, \cite{Soon2019} found an azimuthal variation of the spectral index $\beta$, consistent with a dichotomy between small 
micron-size grains, coupled to the gas, and large millimeter grains strongly concentrated in the gas horseshoe structure, in very 
good agreement with the process of dust trapping in an azimuthal pressure maximum. The first piece of evidence of non-Keplerian gas motions 
in the circumbinary disk was found by \cite{Yen2020} using the C$^{18}$O J=1-0 transition line. This suggests a radial gas pressure bump 
in the north of the disk, similarly to what has been observed across the rings in the HD~163296 and AS~209 systems \citep{Teag2018a, 
Teag2018c}. All these indications of dust trapping, as well as the large size and brightness of the circumbinary disk around HD~142527, 
make of this system the perfect target to look for the existence of a vortex in a protoplanetary disk.

\begin{figure*}[t]
  \includegraphics[angle=0,width=\textwidth]{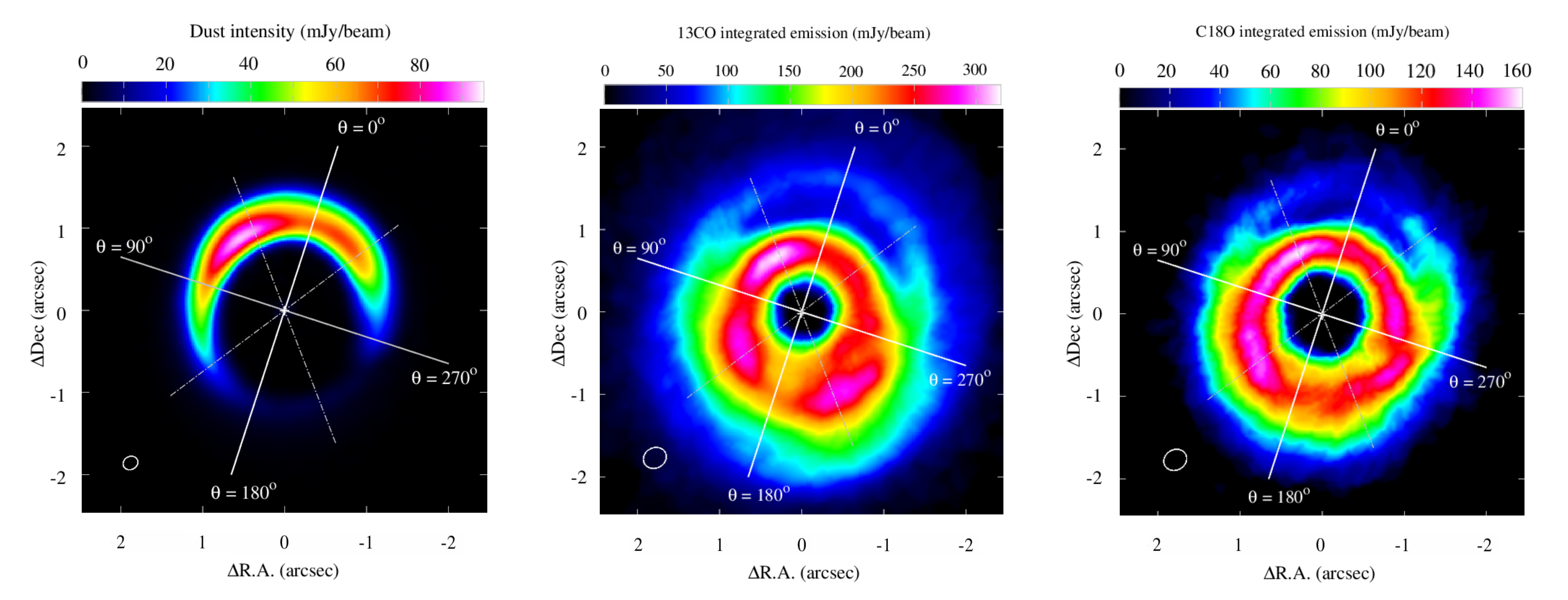}\\
  \caption{From left to right: Dust emission, $^{13}$CO and C$^{18}$O J = 3-2 integrated line emission after continuum subtraction 
  around the binary system HD~142527. The spatial resolution is indicated by the white ellipse at the bottom-left corner of the panels. 
  Its value is 0.''207 $\times$ 0.''178 for the dust emission, obtained using super-uniform weighting, and 0.''31 $\times$ 0.''27 
  for the gas emission, obtained with the Briggs parameter fixed to 0.5. The rms noise is 90 $\mu$Jy beam$^{-1}$ for the dust emission, 
  5.9 mJy beam$^{-1}$ per channel in $^{13}$CO and 7.8 mJy beam$^{-1}$ in C$^{18}$O. The integrated images for $^{13}$CO and C$^{18}$O 
  only take channels with a signal larger than 5 $\sigma$ into account.}
  \label{Fig:dust-CO}
\end{figure*}

Here we present our data on the circumbinary disk around HD~142527 and perform a detailed analysis of the gas kinematics 
(Sect. \ref{sec:Obs}). By comparison with models, we then show that the measured deviations are consistent with the presence of a large 
vortex (Sect. \ref{sec:analysis}). In Sect. \ref{sec:bias}, we examine biases in the measurement of the gas velocity due to beam 
smearing and inspect how they may compare with the current observations around the horseshoe structure. In Sect. \ref{sec:disc}, 
we discuss the vortex scenario and develop strategies to distinguish true velocity signals from artifacts, both in HD~142527 and around 
kinks, spirals, and ring-like structures present in other disks. In sect. \ref{sec:concl}, we summarize our findings.


\section{Observations}
\label{sec:Obs}

\subsection{Morphology of the dust and gas emission}
\label{sec:morph}

This work is based on the analysis of the observations of the HD142527 disk obtained with ALMA (project 2012.1.00725.S), and already published 
in \cite{Boeh2017}. We refer the reader to this paper for a description of the 
observational setup and data calibration. For the sake of completeness, we show in Fig.~\ref{Fig:dust-CO} images of the dust continuum, 
as well as the spectrally integrated intensity (moment 0) maps of the $^{13}$CO and C$^{18}$O J=3-2 line emission obtained in ALMA band 7 
($\sim345$ GHz). For each pixel in the integrated emission maps, we only kept the channels with a signal-to-noise higher than 5 $\sigma$. 
This procedure slightly underestimates the total emission but yields a better signal to noise, especially in the outer regions where 
line emission only comes from a few channels. 

The major axis of the disk has a position angle (PA) of -19$^{\circ}$ relative to celestial north. The disk is rotating in the clockwise 
direction and has an inclination of 27$^{\circ}$ \citep{Fuka2013}, with the far side toward the east (i. e., on the left in the 
figures). The azimuthal angle $\theta$ starts from the major axis (north side) and is counted positively counterclockwise. It is 
measured in the disk plane and therefore slightly differs, due to the disk inclination, from the usual PA measured in the 
image plane. A few azimuthal angles are indicated in Fig.~\ref{Fig:dust-CO}, spaced by 45$^{\circ}$. 

The dust emission around the binary system HD~142527 has a horseshoe structure with a maximum in emission of 93 mJy beam$^{-1}$, or 
34.0 K in brightness temperature taking the inverse of the Planck function. It is located at $\theta$ = 52$^{\circ}$ and at a radius 
of $\sim$ 166 au, taking a distance of 157 pc for the system. The dust emission is not azimuthally symmetric around the maximum of 
emission but features a secondary maximum in the clockwise direction at $\theta$ = -20$^\circ$ with a value of 71 mJy beam$^{-1}$ 
(or 27.5 K). This double-peaked structure may trace a similar structure in dust surface density or may only come from a local 
decrease in temperature due to the shadow, seen in infrared thermal emission and scattered light \citep{Verh2011, Mari2015}, cast
by the inner disk surrounding the main star on the circumbinary disk.

\begin{figure*}[ht]
  \includegraphics[angle=0,width=\textwidth]{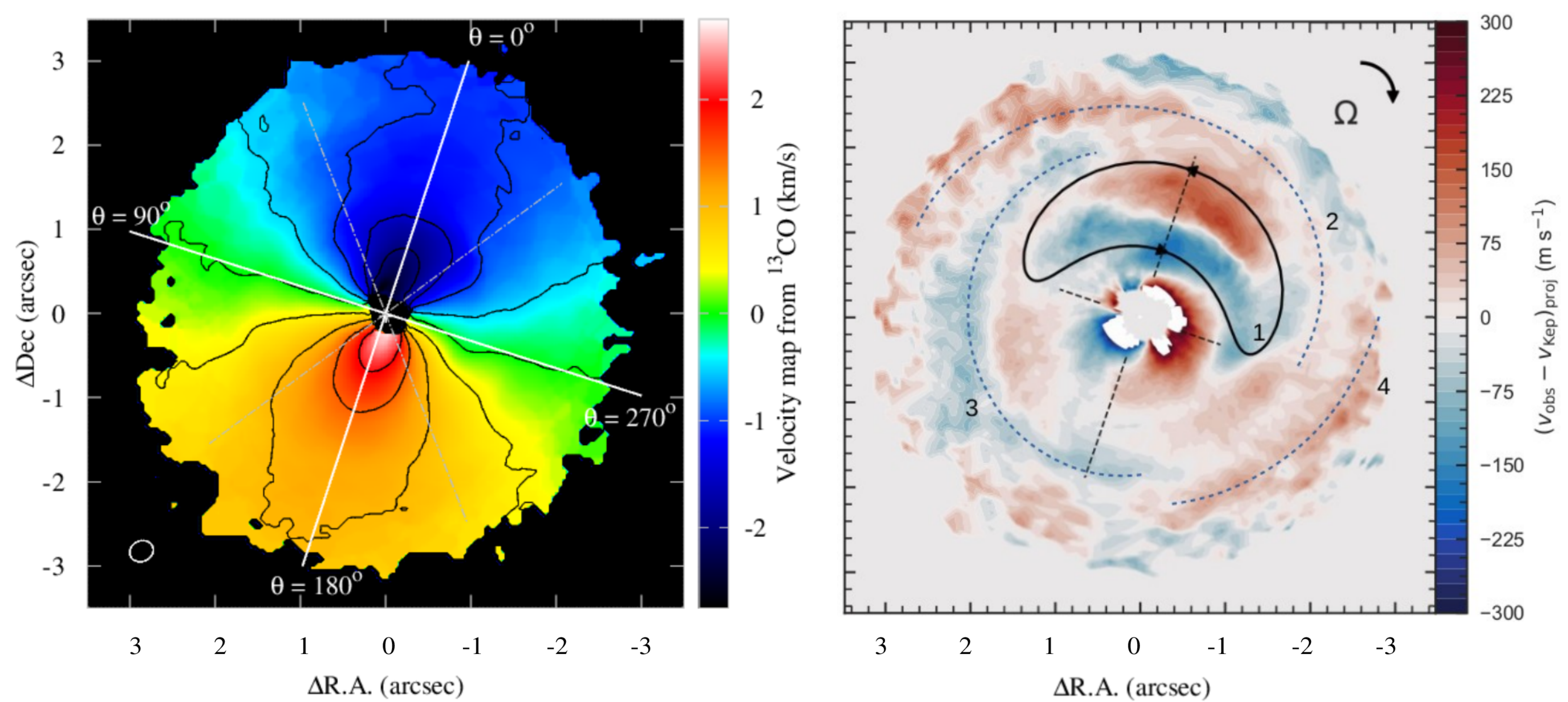}\\
  \caption{Gas velocity of the circumbinary disk. Left: $^{13}$CO J=3-2 velocity map performed using the intensity weighted method, and 
  subtracted by the systemic velocity 
  of 3.73 km s$^{-1}$. Contours are displayed from -2.75 to 2.75 km s$^{-1}$ and are spaced by 0.5 km s$^{-1}$. Right: Blue- and 
  redshifted velocities (indicated by blue and red colors) along the line-of-sight compared to the Keplerian rotation, as expressed by eq. 
  \ref{eq:Kepl}. A polar ellipse represented by a black solid line denoted (1) indicates the potential presence of a anticyclonic 
  vortex, and three spiral-like structures in black dotted-line denoted (2), (3), and (4) are superimposed.}
  \label{Fig:vel-qual}
\end{figure*}

Another tracer of the disk density is the gas through the $^{13}$CO and C$^{18}$O emission lines. Both isotopologues are detected inside the 
disk cavity, and out to large radii, contrary to the millimeter dust emission that is concentrated in a smaller radial range due to radial 
drift and dust trapping \citep{Muto2015, Boeh2017}. The apparent depletion of the gas emission in the north of the disk, at about 200 au 
from the central star, does not imply a local decrease in the gas density and/or temperature, but is an artifact mainly produced by the 
continuum subtraction method. This process overestimates the amount of dust emission to remove as it does not take into account 
that dust emission can be absorbed at the molecular line frequency \citep{Boeh2017, Weav2018}. The $^{13}$CO J=3-2 and C$^{18}$O 
optical depths are estimated to be about 10-15 and $\sim$ 2-3 at the north side of the circumbinary disk. In our observations, the $^{13}$CO 
and C$^{18}$O emission decrease in the horseshoe structure by about 50\% and 60\% at the center of the line after continuum subtraction.

Additionally, part of the $^{13}$CO and C$^{18}$O emission from the back side of the disk can be absorbed by dust particles located in 
the midplane \citep{Isel2018, Rab2020}. This process can theoretically delete up to half of the emission if the line is optically thin and the 
dust highly optically thick. It can have a similar decrease for optically thick molecules if the emission from the front and back molecular 
layers consist of two different lines. This happens when the disk is sufficiently inclined, with gas emission layers at high altitude, such 
that two different radii, and then two different velocities, are probed along the line-of-sight \citep{Teag2018c}. The absorption of the 
line emission by the dust is, however, probably not the dominant process around the horseshoe structure, given the magnitude of the continuum 
subtraction effect, and the fact that the considered transition lines are optically thick and the disk faintly inclined.

\subsection{Overview of the gas kinematics}
\label{sec:kinematics}

The velocity map from the $^{13}$CO J = 3-2 is shown on the left panel of Fig.~\ref{Fig:vel-qual}. The velocity was obtained using 
the intensity weighted method, or moment 1 of the velocity in CASA \citep{McMu2007}. This is the method we favored in the present work 
as the peak emission method appeared more sensitive to the rms noise within our data. A comparison of both methods is, however, given in 
Appendix~\ref{sec:mom9} and visible in Fig.~\ref{Fig:vel-mom9}. A careful study of the velocity can be performed by comparing the gas 
kinematics with the projected Keplerian velocity along the line-of-sight\::
\begin{equation}
  V_{\mathrm{proj}}(r, \theta) = (G M_\star/ r)^{0.5} \sin(i) \cos(\theta),
  \label{eq:Kepl}
\end{equation} 
where $G$ is the Newtonian gravitational constant, $M_\star$ the stellar mass, $i$ the disk inclination, $r$ the orbital radius, 
and $\theta$ the azimuthal angle. We discarded the inner 100 au of the system because our current spatial resolution prevent us to 
sample precisely the disk velocity in the inner region of the disk, but also to avoid known perturbations of the gas flow due to 
the binary \citep{Casa2015}. We used a geometrically thin disk for the fit because we could not precisely constrain the gas scale height, 
while including it in the global fit did not change the value of the other parameters by more than 1$\sigma$. This is explained by the 
low inclination of the disk and the moderate spatial resolution of the data, but also by the intensity weighted method that probes both the 
front and back molecular layers of the disk. \cite{Teag2018c} also showed that assuming a gas scale height only created very small 
differences in the measured gas velocity for the disk around AS~209 due to its moderate inclination of 35$^{\circ}$, even with the peak 
emission method, the $^{12}$CO molecule that emits at a high altitude, and a spatial resolution of 0.2\arcsec.

The 2D map of the velocity deviations from the Keplerian profile is shown in the right panel of Fig.~\ref{Fig:vel-qual} while radial 
cuts for different azimuthal angles are presented in Fig.~\ref{Fig:vel-azi}. We used the software Eddy described in \cite{Teag2019a} 
which is based on an MCMC method, and takes into account the signal-to-noise in each pixel of the image. We fixed the disk inclination 
to 27$^{\circ}$ and fit the mass of the binary, the systemic velocity of the system, the PA of the disk, and the 
center of rotation. Large discrepancies were found depending on the radial distance range considered. For R between 100 and 200 au, 
we obtain a binary mass of 2.53 $\pm$ 0.04 M$_\odot$, where for R $>$ 200 au, we obtain a binary mass of 2.29 $\pm$ 0.07 M$_\odot$. 
These variations are due to the radial Doppler shifted structure denoted 1 at the north of the disk in the right panel of 
Fig.~\ref{Fig:vel-qual}, covering a large area, and which presents super-Keplerian velocities at the inner side of its structure 
and sub-Keplerian velocities at the outer side.

  Performing a fit throughout the circumbinary disk favors the regions at small radii due to a significantly higher signal-to-noise, 
yielding a binary mass of 2.48 $\pm$ 0.03 M$_\odot$. Therefore, to avoid biasing toward any specific radii, we proceeded in two steps. 
First, we used the MCMC code Eddy well outside from the cavity (i. e., R $\geq$ 150 au) to avoid any potential perturbations by the 
companion \citep{Casa2015} and kept the value of the center of rotation ($x_0$,$y_0$) $=$ (15h 56min 41.872s -42d 19min 23.694s) 
located 40 mas south and 20 mas west of the phase center, and of the PA $=$ 161.1$^{\circ}$ $\pm$ 0.4$^{\circ}$  of the disk. 
Second, we used a chi-square method along the minor and major axes of the disk to fit the mass and the systemic velocity V$_{lsr}$ 
of the system between 100 and 400 au, without taking radial variations of the signal-to-noise ratio into account, and obtain a 
binary mass of 2.36 $\pm$ 0.07 M$_\odot$, and a V$_{lsr}$ of 3730 $\pm$ 20 m s$^{-1}$. 

 Aside from the inner region that we cannot precisely sample and which displays non-Keplerian flows due to the presence of the 
binary, four kinematic features are shown in the right panel of Fig.~\ref{Fig:vel-qual}. They are present independently of the 
exact fitting procedure used, and consist in 1) a large radial Doppler-shifted 
structure in the north of the disk with projected velocities up to 160 m s$^{-1}$, roughly colocated with the dust crescent and 
possibly tracing an anticyclonic vortex (cf section \ref{sec:analysis}), 2) a redshifted arc of $\sim$ 180$^{\circ}$ in azimuth 
at the north of the disk, which may be related to the spiral S1 observed in near-IR scattered light, in $^{12}$CO and in $^{13}$CO 
by \cite{Fuka2013}, \cite{Chri2014}, and \cite{Garg2020}, 3) a blueshifted arc of $\sim$ 180$^{\circ}$ in azimuth at the 
east of the disk, which may be the dynamical counterpart of the spiral S4 recently observed by \cite{Garg2020}, and 4) a smaller 
redshifted arc at the southwestern side of the disk. All these arcs and spirals have maximum projected velocity deviations 
on the order of 50 m s$^{-1}$. This supports the idea that the outer region of the disk may present a radial succession of 
spiral and inter-spiral structures, each of them with deviations to the Keplerian rotation. Such spirals have also been recently observed 
at larger radii by \cite{Garg2020} using the $^{12}$CO J=2-1 emission line.


\section{A vortex-like kinematic signal around the dust concentration}
\label{sec:analysis}

\subsection{Preliminary analysis}
\label{sec:pre-analysis}

\begin{figure*}[ht]
  \includegraphics[angle=0,width=\textwidth]{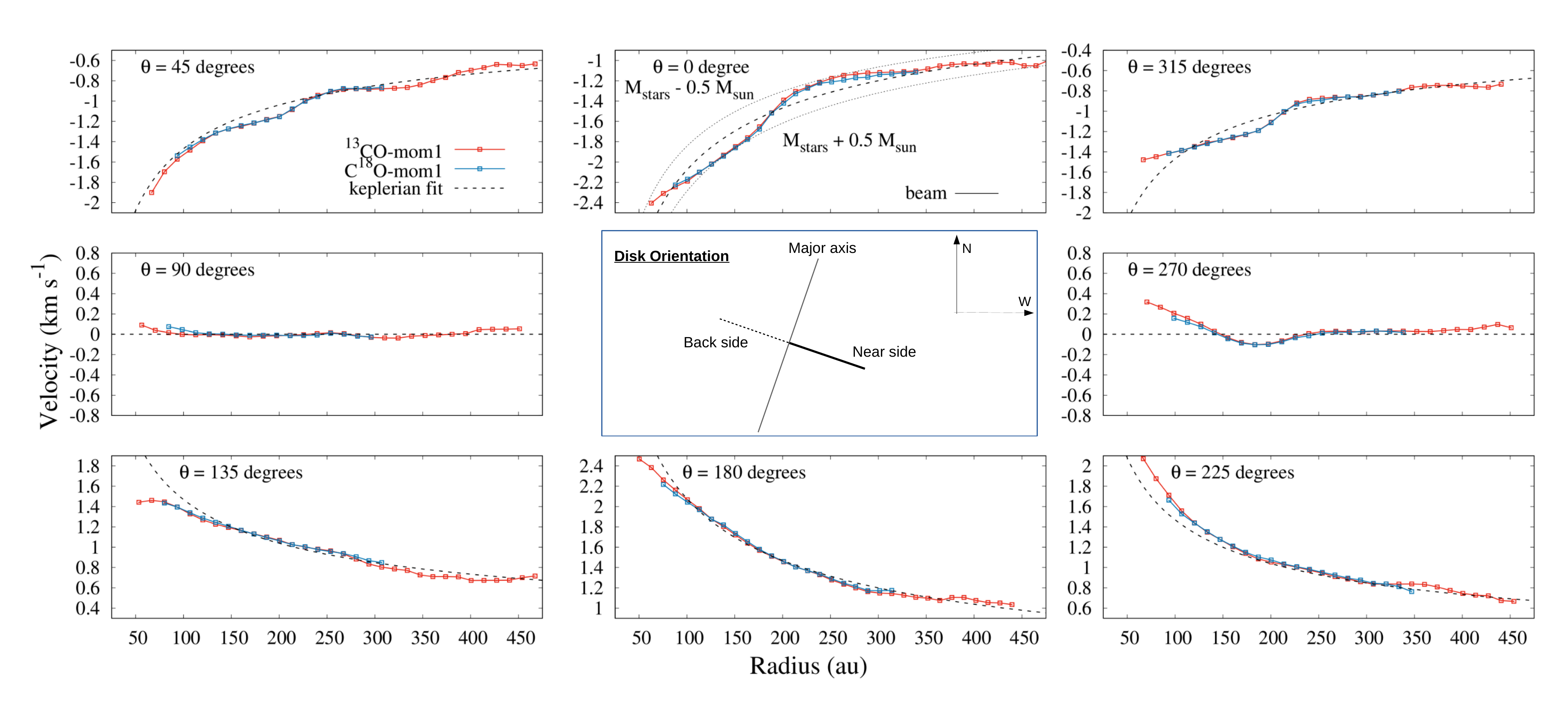}\\
  \caption{Radial profiles of the velocity measured in $^{13}$CO and C$^{18}$O J = 3-2 subtracted by the systemic velocity displayed for 
  various azimuthal angles $\theta$, spaced by 45$^{\circ}$. The dashed line represents the Keplerian prescription following equation 
  \ref{eq:Kepl} for a 2.36 $\mathrm{M}_\odot$ binary system. The two dotted lines in the top central panel indicate the Keplerian velocity 
  for a binary mass of 2.86 $\mathrm{M}_\odot$ and 1.86 $\mathrm{M}_\odot$.}
  \label{Fig:vel-azi}
\end{figure*}

In this study, we focus on the radial Doppler-shifted feature, denoted (1) in the right panel of Fig.~\ref{Fig:vel-qual}, and whose 
projected deviations to the Keplerian rotation reach 160 m s $^{-1}$, compared to $\sim$ 50 m s$^{-1}$ along the spiral arcs. Radial 
cuts of the velocity deviations at different azimuthal angles 
are shown in Fig.~\ref{Fig:vel-azi}. The velocities measured using the $^{13}$CO and C$^{18}$O lines give very similar values, 
with differences no larger than 20-30 m s$^{-1}$ between $\sim$ 100 and $\sim$ 350 au. This suggests that both isotopologues emit 
from a similar altitude and that the precision on the velocity measurement is only of a few tens of m s$^{-1}$, well below the 
channel width of $\sim$ 110 m s$^{-1}$ in the observations. 

The gas velocity is correctly matched by a Keplerian profile, represented by the black dashed line, on the Southern side of the disk. 
On the contrary, the gas velocity along the northern major axis presents a clear distinct S-profile that cannot be approximated by a 
single power law. The sensitivity to azimuthal motions is maximized along the major axis due to projection effects. On the 
contrary, we are blind to radial motions because they are perpendicular to the line-of-sight. We may also observe vertical motions 
as their projection along the line-of-sight is independent of the azimuthal angle. However, at first order, they can probably be 
neglected in the radial Doppler-shifted structure 
as 3D simulations have shown that vertical motions around vortices in steady state are negligible \citep{Lin2012, Rich2013} and 
while spirals can present vertical motions, their projected deviations are not higher than 50 m s$^{-1}$ in the rest of the disk 
and with a different morphology.

Taking into account the clockwise rotation of the disk with the far side being in the east (i.e., on the left), the gas 
rotates at a super-Keplerian speed at radii between $\sim$ 80 and 185 au, and at a sub-Keplerian speed at radii between $\sim$ 
185 and 300 au, confirming the deviations observed in \cite{Yen2020}. Projected velocities are of about $\pm$ 0.16 km s$^{-1}$, 
or of 0.35 km s$^{-1}$ after deprojection, about 10\% of the background Keplerian rotation. The transition radius between these 
two regimes is at 185 au and corresponds to the distance at which the dust is radially concentrated in the horseshoe structure 
\citep{Boeh2017, Soon2019, Yen2020}. 

These azimuthal velocity deviations may only come from radial gas pressure gradients due to the outer ring-like structure \citep{Teag2018a}. 
It may also trace the azimuthal deviations coming from a vortex centered near this position \citep{HuanP2018, Robe2020}. The vortex 
presence would imply radial deviations to the Keplerian rotation. At the adjacent angles $\theta$ 
= 45$^\circ$ and 315$^\circ$, we are equally sensitive to radial and azimuthal motions. The interpretation of the projected velocities along 
the line-of-sight is therefore not as direct but the S-shape previously seen along the major axis is, however, still visible, suggesting that 
the azimuthal behavior of the gas described at $\theta$ $=$ 0$^\circ$ is still present. In addition, velocities are slightly blueshifted at both 
angles indicating, at $\theta$ = 45$^\circ$, that the gas is flowing inward in the radial direction while, at $\theta$ = 315$^\circ$, 
gas is flowing radially outward. Finally, along the minor axis, we are only sensitive to radial velocity deviations. At the near side, 
outward gas motion is measured at R $\sim$ 185 au while at the far side, at $\theta$ $=$ 90$^\circ$, the gas motion does not display 
any radial deviations. 

The preliminary analysis of the velocity deviations in the north of the disk shows then that, at first glance, they are compatible with 
the presence of a large vortex located at a radius of 185 au, centered between 315$^\circ$ and 45$^\circ$, and extending azimuthally over 
about half of the circumbinary disk. A sketched view of this kinematic structure denoted (1) is shown in the right panel of 
Fig.~\ref{Fig:vel-qual}. Alternatives to this scenario are also discussed in Sect.~\ref{sec:bias}.




\subsection{Vortex model}
\label{sec:vortex-pres}

To better compare the deviations to the Keplerian rotation with the kinematic signature of a vortex, a model is required. Simulations 
by \cite{HuanP2018} and \cite{Robe2020} have shown that the vortex size and its associated velocity can largely vary as a function of the process 
responsible for its origin (massive planet, binary, dead zones), the underlying gas surface density, or the disk viscosity. Our goal here is not 
to perform a full simulation able to reproduce the vortex-like kinematic signal but, on the contrary, to use a simple model to constrain 
the main parameters of a possible vortex, such as its position, size, aspect ratio, and velocity. 

\begin{figure}[ht]
  \includegraphics[angle=0,width=0.5\textwidth]{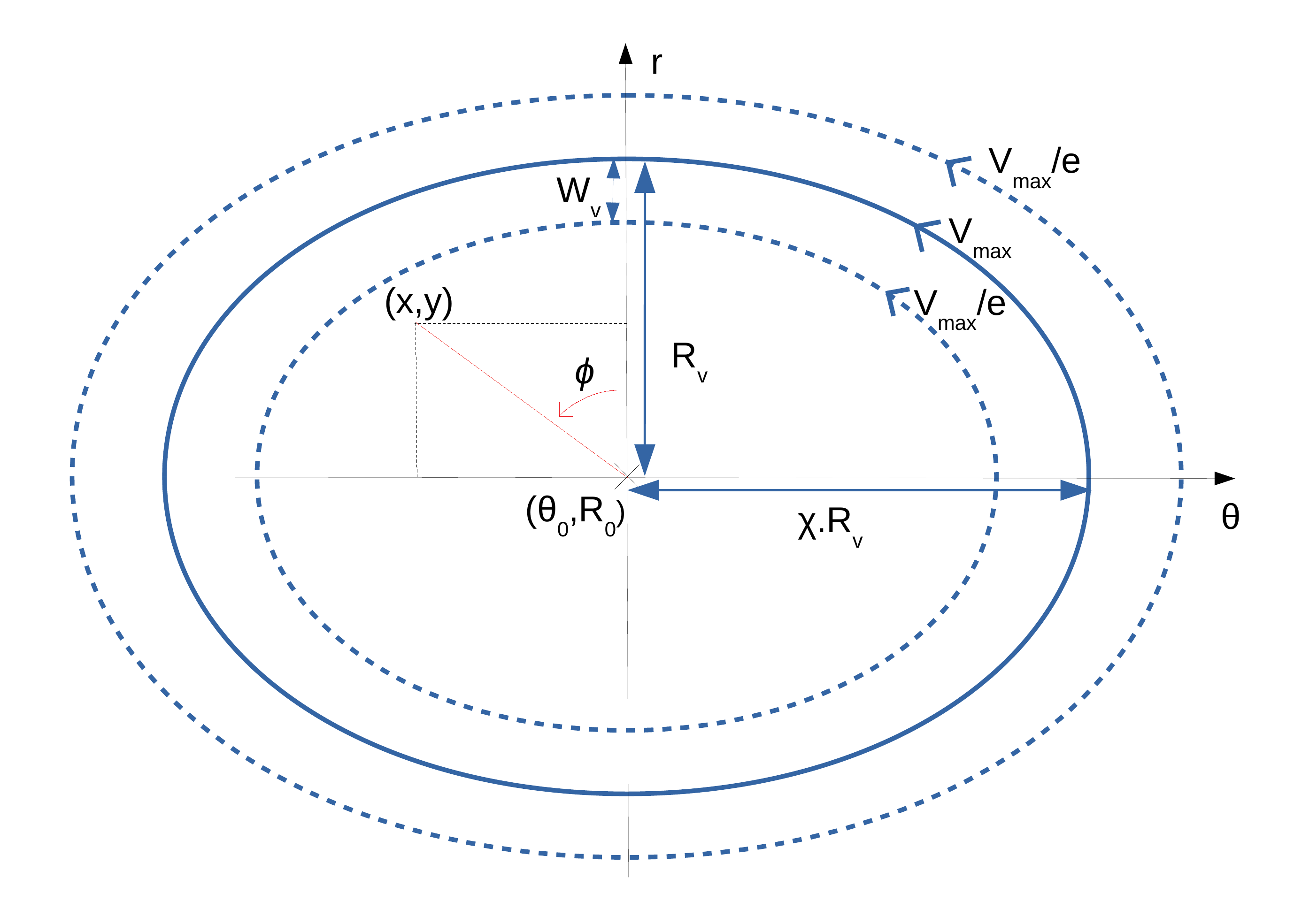}\\
  \caption{Sketch describing the vortex and its parameters. Such a vortex may be located on the northern side of the disk, at a similar 
  localization than the polar ellipse denoted (1) and drawn in the right panel of Fig.~\ref{Fig:vel-qual}.}
  \label{Fig:vortex-shema}
\end{figure}

A sketch describing the vortex parameters is shown in Fig.~\ref{Fig:vortex-shema}. In the polar coordinates of the disk, the flow of the gas 
due to a vortex is often described with elliptic streamlines of constant velocity characterized by a central position ($R_\mathrm{0}$, 
$\mathrm{\theta}_0$) and an aspect ratio $\mathrm{\chi}$ $=$ $b/a$, with $b$ the major axis in the azimuthal direction and $a$ the minor axis 
in the radial direction \citep{Kida1981, Good1987, Chav2000, Surv2015}. The flow of the gas is null at the vortex center ($R_\mathrm{0}$, 
$\mathrm{\theta}_0$), then it increases with distance to that center, up the velocity $V_{\mathrm{max}}$ before decreasing again to a null 
value. The exact vortex velocity profile cannot be measured from our current data, and we assume it has a Gaussian profile. Along the radial 
axis, we note $R_\mathrm{v}$ the distance from the vortex center at which the velocity reaches its maximum, $V_\mathrm{max}$, and 
$w_\mathrm{v}$ the half-width of the Gaussian. Azimuthally, the maximum in velocity is reached at the distance of $\mathrm{\chi}R_\mathrm{v}$ 
from the vortex center. 


For simplicity, we note in the following the distance to the vortex center in Cartesian coordinates ($x$,$y$) 
with $x$ = $R_\mathrm{0}$ $ \times$ ($\theta$ - $\theta_0$) and $y$ = $r$ - $R_\mathrm{0}$. Along the radial axis, the absolute 
value of the vortex velocity is then:
\begin{equation}
|V(0,y)| = V_\mathrm{max} \enspace \exp^{-\left ( \frac{|y|-R_\mathrm{v}}{w_\mathrm{v}} \right ) ^{2} }.
\end{equation}
Thereafter, we can obtain the vortex velocity at any point (x,y) of the disk by finding the ellipse, of constant velocity, which crosses 
both this position (x,y) and the radial axis along the vortex eye. The elliptic streamlines of the vortex can be compared to circles inclined 
along the azimuthal axis. At the position ($x$,$y$) and angle $\phi$ of the vortex, we are at the position ($x$, $\chi_y$) and
at the angle $\phi^\prime$ of such a circle. $\phi^\prime$ is defined between the radial axis and the current position, and counted 
counterclockwise such that:
\begin{equation}
\cos(\phi^\prime) =  \frac{\chi y}{\sqrt{x^2+(\chi y)^2}}.   
\end{equation} 
The ellipse going through the position ($x$,$y$) will then cross the radial axis of the vortex at the position $y$/cos($\phi^\prime$), 
with the absolute velocity:
\begin{equation}
|V(x,y)| = |V(0, y/\cos(\phi^\prime))| =  V_\mathrm{max} \enspace \exp^{-\left ( \frac{|y/\cos(\phi^\prime)|-R_\mathrm{v}}{w_\mathrm{v}} 
            \right ) ^{2} }.
\end{equation}
The final step is to calculate the velocity along the line-of-sight; the only observable quantity. This requires us to decompose 
the vortex velocity into its radial ($V_\mathrm{r}$) and azimuthal ($V_\mathrm{\theta}$) components. Using the inclination $i$ of 
the disk, the disk azimuthal angle $\theta$, and the angle $\phi^\prime$ in the vortex reference frame, the projected velocity along the 
line-of-sight is given by:
\begin{eqnarray}
   V_{\mathrm{proj}} & = & \left[ V_r ~ \sin(\theta) + V_{\theta} ~ \cos(\theta) \right] \sin(i), \nonumber \\
                     & = & |V|  \left[ - \sin(\phi^\prime) \sin(\theta) + \cos(\phi^\prime) \cos(\theta)  \right] \sin(i), \nonumber   \\
                     & = & |V| ~ \cos(\theta + \phi^\prime) ~ \sin(i).
\end{eqnarray}  
 
\subsection{Comparison of the observations with the vortex model}

\begin{figure*}[h]
  \includegraphics[angle=0,width=\textwidth]{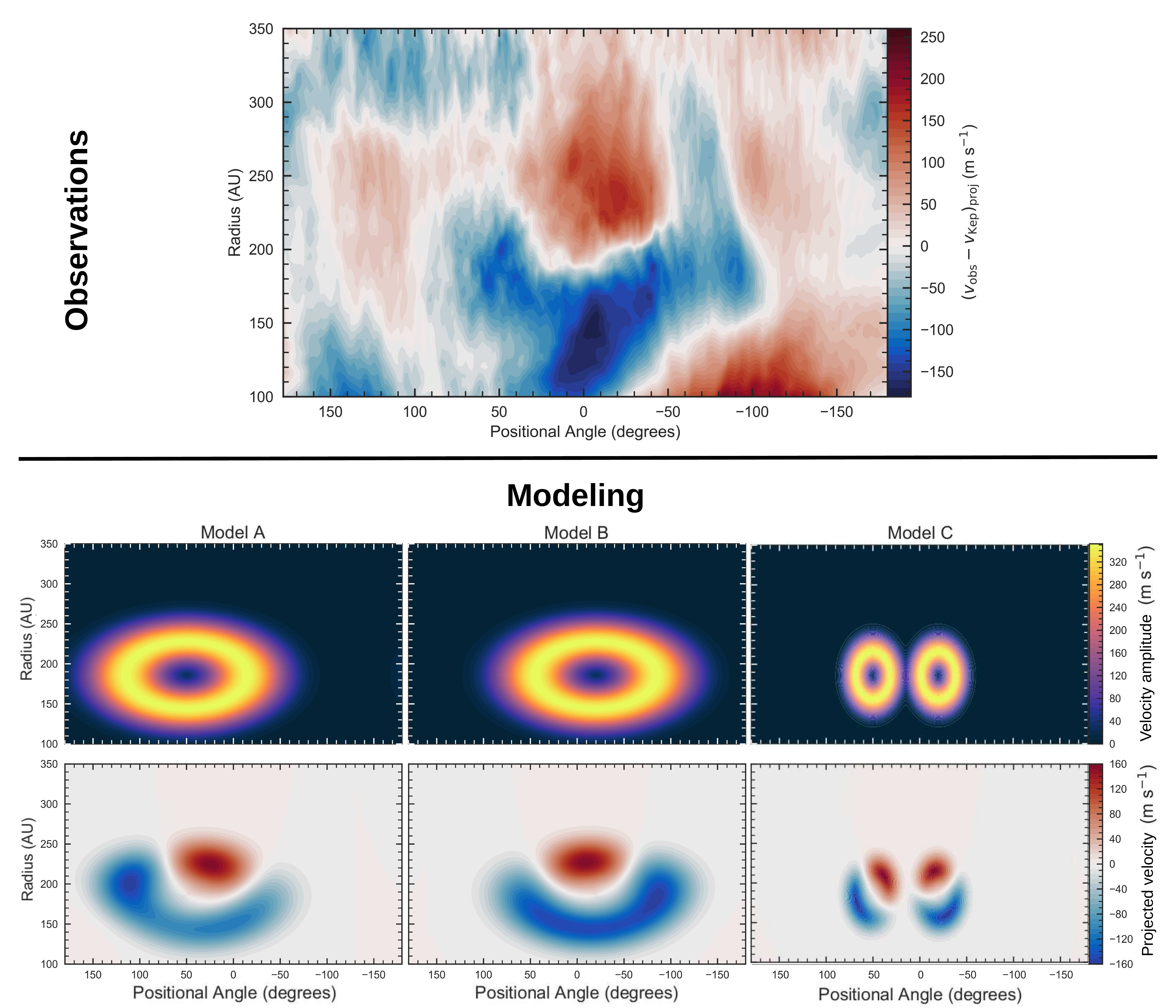}\\
  \caption{Comparison of the velocity deviations measured in the HD~142527 system with vortex models. Top: Difference between the $^{13}$CO J=3-2 
  velocity measured from our observations and the Keplerian prescription in a 2D-map 
  ($R$,$\theta$). Blue and red colors indicate blue- and redshifted velocities along the line-of-sight compared to the Keplerian rotation. 
  $\theta$ = 0$^\circ$ represents the north of the major axis. Bottom: Three vortex prescriptions with their corresponding velocity 
  amplitude in the plane of the disk (top), rotating clockwise around the vortex center on elliptical streamlines, and the projected 
  velocity along the line-of-sight (bottom). Model A corresponds to a large vortex centered at the position of the continuum intensity 
  maximum and Model B an equivalent vortex centered on the secondary maximum. Model C presents the kinematic signature for two 
  smaller vortices at these positions.}
  \label{Fig:vortex-mod}
\end{figure*}

This comparison is focused in the circumbinary disk at a radius between 100 and 350 au, where the Doppler shifted structure denoted (1) 
in the right panel of Fig.~\ref{Fig:vel-qual} is present. In the polar 2D map ($r$, $\theta$) shown in the top panel of
Fig.~\ref{Fig:vortex-mod}, the observed kinematic feature consists in a redshifted clump located between -50$^{\circ}$ to 50$^{\circ}$ 
and at radii between 180 and 300 au surrounded by a large blueshifted arc located at PA between -100$^{\circ}$ and 100$^{\circ}$ and at 
radii between 120 and 200 au. This velocity structure has maximum projected kinematic deviations of $\sim$ 150 m s$^{-1}$ and is 
centered at $\sim$ 185 au, at a radius similar to the dust and gas maximum surface densities inside the dust crescent \citep{Muto2015, 
Boeh2017}.

In the bottom panels of Fig.~\ref{Fig:vortex-mod}, we applied our vortex prescription to three different possibilities, whose 
parameters are given in Table~\ref{tab:vortices}. In models A and B, we made the hypothesis 
that the dust concentration is due to a single large vortex. Its size corresponds to the extension of the kinematic structure, with R$_v$ 
$=$ 42 au, W$_v$ $=$ 26 au, and an azimuthal-to-radial aspect ratio of $\sim$ 5. The maximum velocity reached by the vortex is set to 
350 m s$^{-1}$, a value constrained by the deprojection of the observed velocity. This value is comparable to the local sound 
speed for a temperature of 30-40 K at the horseshoe position using the formula $\sqrt{k_b T /(\mu m_h)}$, with $k_b$ the Boltzmann constant, 
$\mu$ = 2.3 the mean molecular weight, and $m_h$ the hydrogen mass. The only difference between Models A and B is 
the azimuthal position with a vortex centered at the position of the dust emission maximum ($r =$ 185 au, $\theta =$ 50$^{\circ}$) in 
model A and at the position of the secondary maximum ($r =$ 185 au, $\theta =$ -20$^{\circ}$) in model B. We further tested a third model 
(Model C) where two smaller vortices are located at the two maxima in the dust emission.

From these three models, it appears that our comparison favors the option of a large and single vortex, located near the secondary 
maximum in dust emission at PA $=$ - 20$^\circ$. The model with two vortices, on the contrary, does not reproduce the morphology of 
the kinematic signal. Furthermore, the two vortices would have a small aspect ratio of $\sim$ 2 to cover the extent of the kinematic 
deviations, and would therefore probably not withstand the elliptic instability \citep{Lesu2009}.


\begin{center}
\begin{table}  
  \begin{tabular}{|p{3.1cm}|c|c|cc|}
    \hline
                            & Model A      & Model B     &  \multicolumn{2}{c|}{Model C} \\    
                            &              &             &       V1      &      V2      \\ \hline                
    R$_0$ (au)              & 185          & 185         &      185      &     185      \\
    $\theta_0$ (degrees)    & 50$^\circ$   & -20$^\circ$ &  50$^\circ$   &  -20$^\circ$ \\ 
    $\chi_A$ (aspect ratio) & 5            & 5           &  2.0          &   2.0        \\
    V$_{\mathrm{max}}$ (m s$^{-1}$)  & 350          & 350         &  350          &   350        \\
    R$_v$     (au)          & 42           & 42          &  35           &   35         \\ 
    w$_v$     (au)          & 26           & 26          &  22           &   22         \\ \hline    
  \end{tabular}
  \caption{Parameters used in the three models to describe the vortex properties.}     
  \label{tab:vortices}
\end{table}   
\end{center}

\section{\textbf{Other origins of kinematic deviations around the horseshoe structure}}
\label{sec:bias}

\subsection{Velocity deviations are also observed in pure Keplerian disks}

\begin{figure*}[ht]
  \includegraphics[angle=0,width=\textwidth]{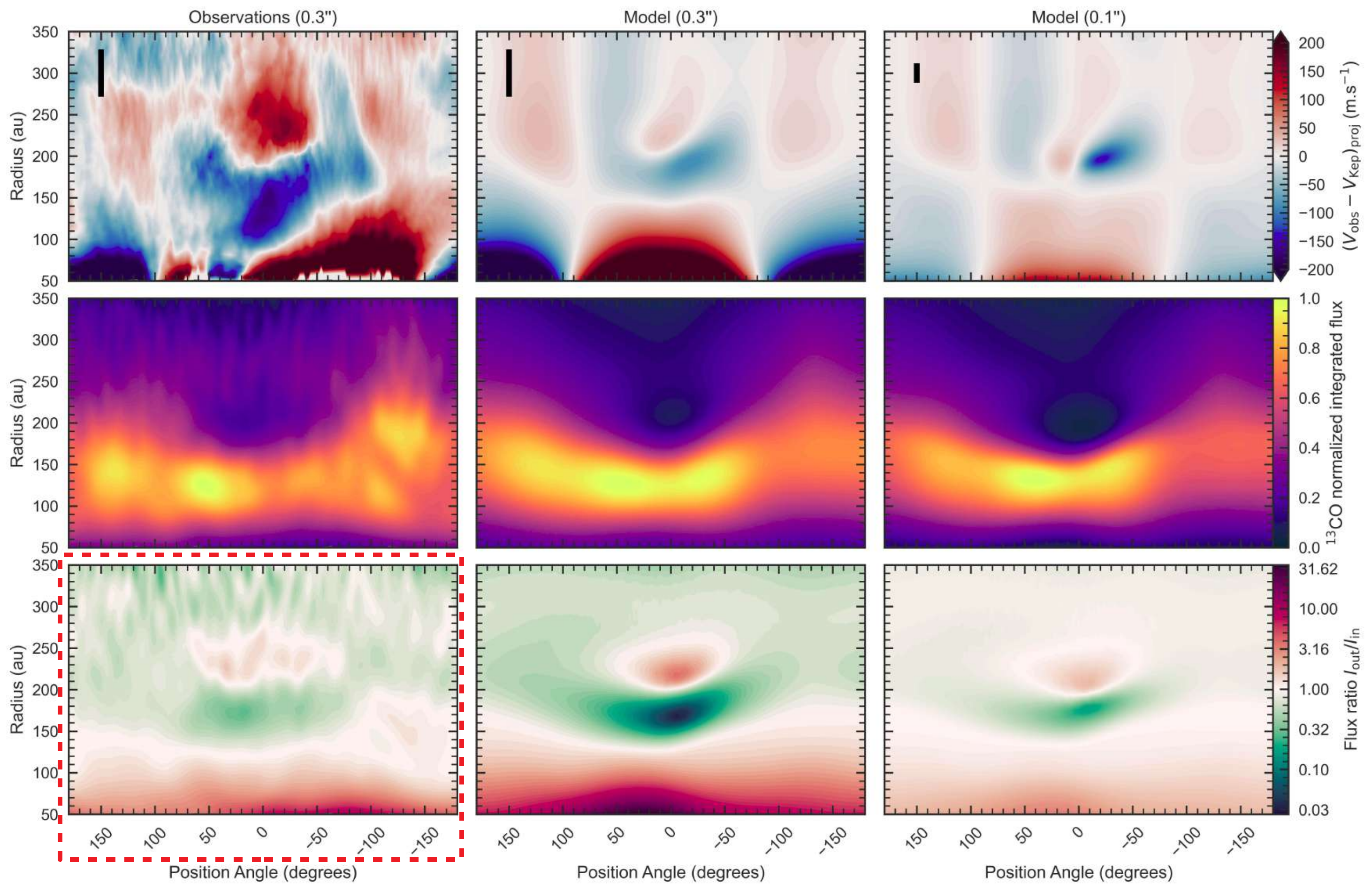}\\
  \caption{Artificial velocity deviations measured in a perfectly Keplerian disk compared to the observations. Top: Differences between 
  the measured $^{13}$CO J=3-2 velocity and the Keplerian prescription in a 2D-map (R,$\theta$). Blue and red colors indicate blue- and 
  redshifted velocities along the line-of-sight compared to the Keplerian rotation. Middle: $^{13}$CO integrated emission. Bottom: Flux 
  ratio I$_{out}$/I$_{in}$ of the emission at the outer and inner edges of the synthesized beam. The left panel is surrounded by a red 
  rectangle because the flux ratio is only an approximation in observations, being measured after beam smoothing. Left column shows the 
  observations. The central and the right columns are the models convolved at a spatial resolution of 0.3 and 0.1\arcsec. The vertical 
  black bar on the top-left corner of the top panels is the averaged radial spatial resolution as a function of the azimuth.}
  \label{Fig:dev-mod}
\end{figure*}

To probe the precision of the velocity measurements, we carried out a 3D model of the circumbinary disk around HD~142527, performed using 
the software RADMC3D \citep{Dull2012} and based on the parameters previously determined in the studies of \cite{Boeh2017}, \cite{Soon2019}, 
and \cite{Yen2020}. A detailed description of the procedure is presented in Appendix \ref{app:model} and images of the model are given in 
Fig.~\ref{Fig:model-image}. In this model, the disk is rotating in pure Keplerian rotation. The resulting signal is then convolved by a 
Gaussian of 0.1\arcsec\ and 0.3\arcsec\ in order to reproduce typical spatial resolutions of molecular lines in ALMA observations. 

We show in the top-center and top-right panels of Fig.~\ref{Fig:dev-mod} that measurable deviations exist even in the case of ideal 
observations (i.e., without noise and at a velocity resolution of 25 m s$^{-1}$). The velocity deviations are of a few percent and can 
be divided into two regions. At R $\leq$ 150-175 au, the velocity deviations appear sub-Keplerian, meaning redshifted in the north of the disk 
(at PA between -90$^\circ$ and 90$^\circ$) and blueshifted in the south of the disk. On the contrary, at R $\geq$ 150-175 au, the gas 
appears mainly super-Keplerian, with an inversion of the blueshifted and redshifted azimuthal locations. The only exception is the area 
near the north of the major axis, which will be detailed in section \ref{sec:dust}.  

In the top-left panel of Fig.~\ref{Fig:dev-mod}, we present the velocity deviations measured in the observations. At a radius larger 
than 150 au, the observed kinematic signal is similar to the velocity pattern seen in the pure Keplerian model at 0.3\arcsec, even if the 
velocity deviations reach a velocity of 150 m s$^{-1}$ in our observations, instead of only 20-80 m s$^{-1}$ in our models. We 
observe the rounded redshifted region at $\sim$ 230 au, close to PA = 0$^{\circ}$, surrounded by the blueshifted area at $\theta$ 
between -90$^{\circ}$ and 90$^{\circ}$, and mainly a redshifted region in the south of the disk. The similarity of these characteristics can 
cast doubts about the correct physical interpretation of the kinematic deviations to the Keplerian rotation and might reveal a bias 
in the velocity measurements to start with. Inside the cavity, at R $\leq$ 100-150 au, the observed velocity deviations diverge from 
our models and probably trace perturbations of the Keplerian flow by the companion, as observed in \cite{Casa2015}. 

\subsection{Line intensity gradients skew measured velocities through beam smearing}
\label{sec-bias1}

Observations are naturally limited by characteristics such as the rms noise, and the spatial and spectral resolutions. Each of them can 
bring systematic biases in the measurements. In the HD~142527 disk models, the two main regions in terms of velocity deviations 
are separated at a radius of about 150 au, corresponding to the distance at which the $^{13}$CO J=3-2 emission peaks. This is visible in 
the middle line in Fig.~\ref{Fig:dev-mod} which shows the spatial distribution of the $^{13}$CO integrated emission. In the inner regions, 
as the $^{13}$CO flux increases with radius, we collect more emission from large radii than from small radii inside the synthesized beam, 
leading to a sub-Keplerian profile. The opposite effect happens in the outer regions where the gas emission mainly decreases with radius, 
leading to a super-Keplerian profile. The spurious velocity deviations in the models are then explained by the finite spatial resolution 
of the observations and the radial gradient of the molecular line emission. This effect, also called ``beam smearing,'' was discussed a 
first time in \cite{Kepp2019} for the kinematic analysis of the PDS 70's cavity (see the appendix A.2).

A useful parameter to understand these artifacts in the velocity measurements is the radial variation of the line emission. For a given 
spatial resolution, this variation can be estimated by the emission ratio I$_\mathrm{out}$/I$_\mathrm{in}$ between the outer and inner edges of 
the synthesized beams. When this ratio is greater than 1, the measured velocity becomes sub-Keplerian, and vice versa. Starting 
from the initial non-convolved image, this ratio can be perfectly known in the models and is shown in the bottom-center and bottom-right 
panels of Fig.~\ref{Fig:dev-mod}. As expected, the emission ratio I$_\mathrm{out}$/I$_\mathrm{in}$ is larger at 0.3\arcsec\ than at 0.1\arcsec\ 
and lead to stronger velocity biases. In real observations, the emission ratio can only be approached because the image received is already 
convolved by the observational beam. We show, however, in the bottom-left panel of Fig.~\ref{Fig:dev-mod} that the general tendency of 
the flux variations can be recovered, even if their magnitude is probably underestimated.

\begin{figure}[ht]
  \includegraphics[angle=-90,width=0.5\textwidth]{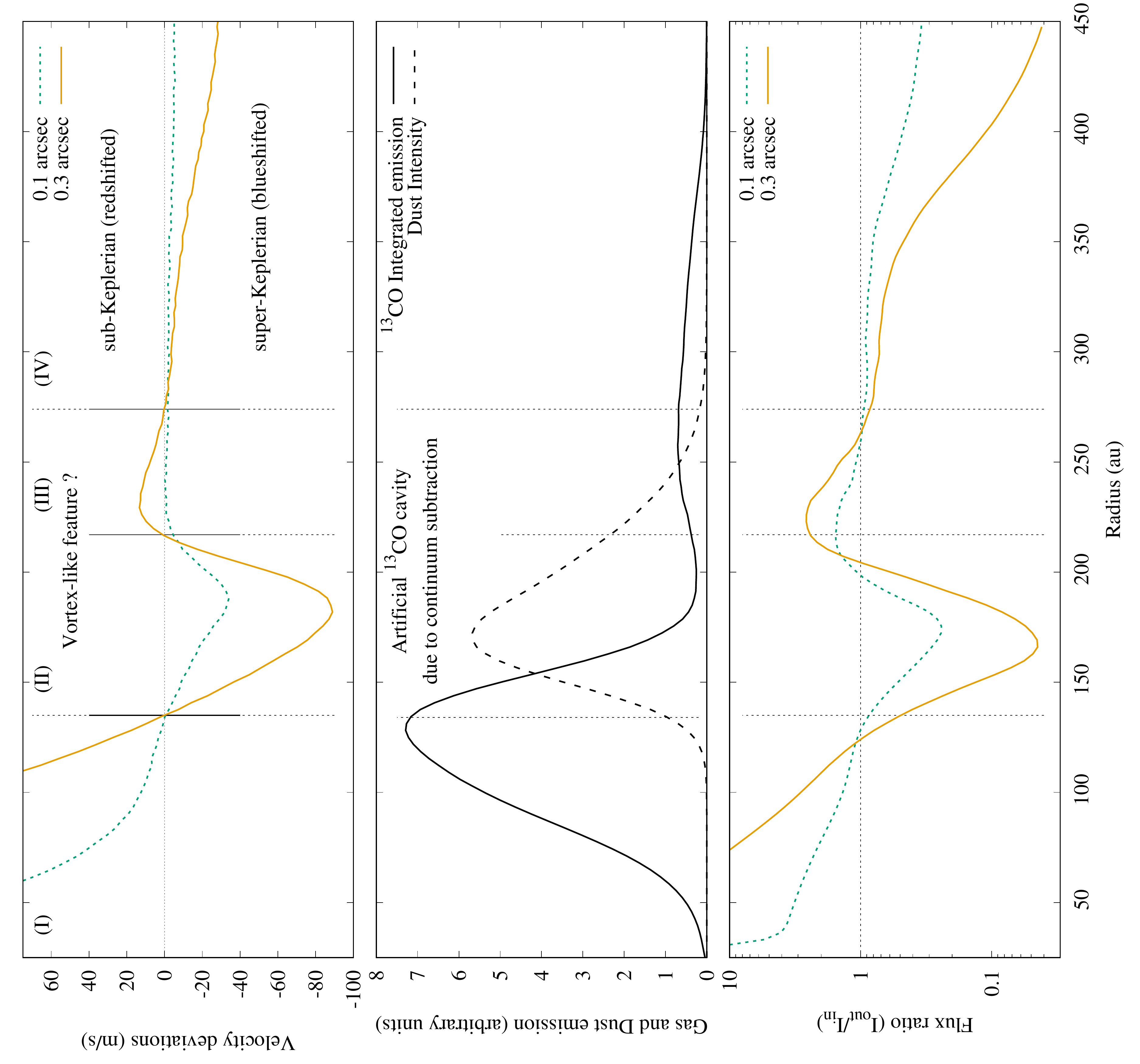}\\
  \caption{Velocity deviations due to the presence of dust along the north of the major axis. Top: Velocity deviations measured in the pure 
  Keplerian model at two different spatial resolutions; Middle: $^{13}$CO integrated emission and dust emission in the original non-convolved 
  model. By definition, the integrated emission is always measured after continuum subtraction. The cavity between 150 au and 260 au is due 
  to the continuum subtraction procedure; Bottom: Flux ratio between the outer and inner edges of the synthesized beam.}
  \label{Fig:dev-quant}
\end{figure}

In general, we find in our models that the flux ratio I$_\mathrm{out}$/I$_\mathrm{in}$ is proportional to the beam size, with ratios at 0.3\arcsec\ 
about three times larger than at 0.1\arcsec, which is an expected result when the spatial variation of the line emission is relatively 
smooth compared to the synthesized beam size. Furthermore, biases are also sensitive to the velocity range $\Delta V$ 
probed by the synthesized beam, which is proportional to the beam size, and increases at smaller distances from the star due to the 
steeper gradient of the Keplerian velocity. This leads to velocity artifacts generally larger at 0.3\arcsec\ compared to 0.1\arcsec\, and 
particularly important in the inner regions of the disk. Along the northern major axis, as shown in the top panel of Fig.~\ref{Fig:dev-quant},
the velocity deviations at 0.1\arcsec\ are about three times smaller than at 0.3\arcsec. Going to a higher spatial resolution will have then 
the double advantage of i) reducing velocity biases and of ii) allowing a better determination of the value and morphology of these biases, 
through a more precise knowledge of the disk structure.


\subsection{Dust also affects the measured gas velocity}
\label{sec:dust}

The circumbinary disk around HD~142527 is not only made of gas but also contains a large concentration of dust particles at the north of 
the disk. The presence of this dust concentration, which is optically thick at 0.9 millimeter with $\tau_{Dust}$ $\sim$ 2-3 \citep{Muto2015, 
Boeh2017}, has an important impact on the measured gas velocity at the north of the disk, as shown in the top panel of Fig.~\ref{Fig:dev-quant}. 
The four radial regions indicated in this figure are delimited by the radii at which the velocity shifts between a sub- and a super-Keplerian 
velocity at the spatial resolution of 0.3\arcsec. Regions I and IV, which contain almost no millimeter dust emission, correspond to the two 
regions described previously in section \ref{sec-bias1}. The $^{13}$CO integrated emission uniformly increases for R $\leq$ 135 au and 
decreases for R $\geq$ 270 au, leading to the sub-Keplerian and to the super-Keplerian profiles. 

The existence of the additional regions II and III, located between $\sim$ 135 and $\sim$ 270 au, is of main interest in our analysis 
as it corresponds to the location of the potential vortex signature in the observations, around the dust horseshoe position. As shown in 
the middle panel of Fig.~\ref{Fig:dev-quant}, velocity biases in the models are related to the artificial and local cavity in the gas emission 
in the north of the disk, visible both in the models and in the observations in Fig.~\ref{Fig:dust-CO}. The gas cavity is colocated 
with the presence of dust emission and mainly arises from the continuum subtraction process, which over-estimates the dust contribution 
to subtract \citep{Boeh2017, Weav2018}, as already discussed in Section~\ref{sec:morph}.

Independently of its origin, real or artificial, the top and bottom panels in Fig.~\ref{Fig:dev-quant} reveal that the measured 
velocity deviations are in good agreement with the flux variations measured after continuum-subtraction. We obtain a super-Keplerian 
profile in Region II (135 $<$ R $<$ 215 au), and a sub-Keplerian profile in Region III (215 $<$ R $<$ 270 au), 
matching the flux ratio smaller and greater than 1, reciprocally. The presence of dust has then an important impact on the measured velocity.

These deviations are not due to the continuum subtraction process, however. The measured velocity does not change before or after 
this process, as shown by \cite{Teag2018c} using the peak emission method, which can be performed indistinctly with or without continuum 
subtraction. Indeed, the emission that we 
subtract is the interpolation of the dust emission from adjacent line-free channels and is essentially a flat spectrum compared to the width 
of the line. For a typical protoplanetary disk located at a distance of $\sim$ 150 pc, the $^{13}$CO J=3-2 line has a spectral width of about 
1-2 MHz at a spatial resolution of 0.1-0.3\arcsec. At a line frequency $\nu$ of $\sim$ 330 GHz, this gives a ratio $\Delta \nu$/$\nu$ of 6 
$\times$ 10$^{-6}$. Therefore, the dust thermal emission, with a spectral index between 2 and 3.7, only varies in amplitude by about 0.001\% 
along the spectral width of the line and does not affect the measured velocity. 

On the contrary, the fact that velocity biases follow flux variations after dust-subtraction indicates that the weight given to each 
regions inside the synthesized beam is not the absolute magnitude of the line but its excess over a ``quasi-flat spectrum,'' which 
can be zero-emission (if there is no dust) or the dust emission (if dust is present). When the gas emission becomes optically thick, 
the combined dust and gas emission at the line frequency is smaller than the sum of the two components taken separately, due to optical 
depth effects. The prominence of the line over the quasi-flat spectrum decreases and the weight given to the regions with more dust 
emission is then reduced for the measurement of the gas velocity. 

\begin{figure*}[ht]
  \includegraphics[angle=0,width=\textwidth]{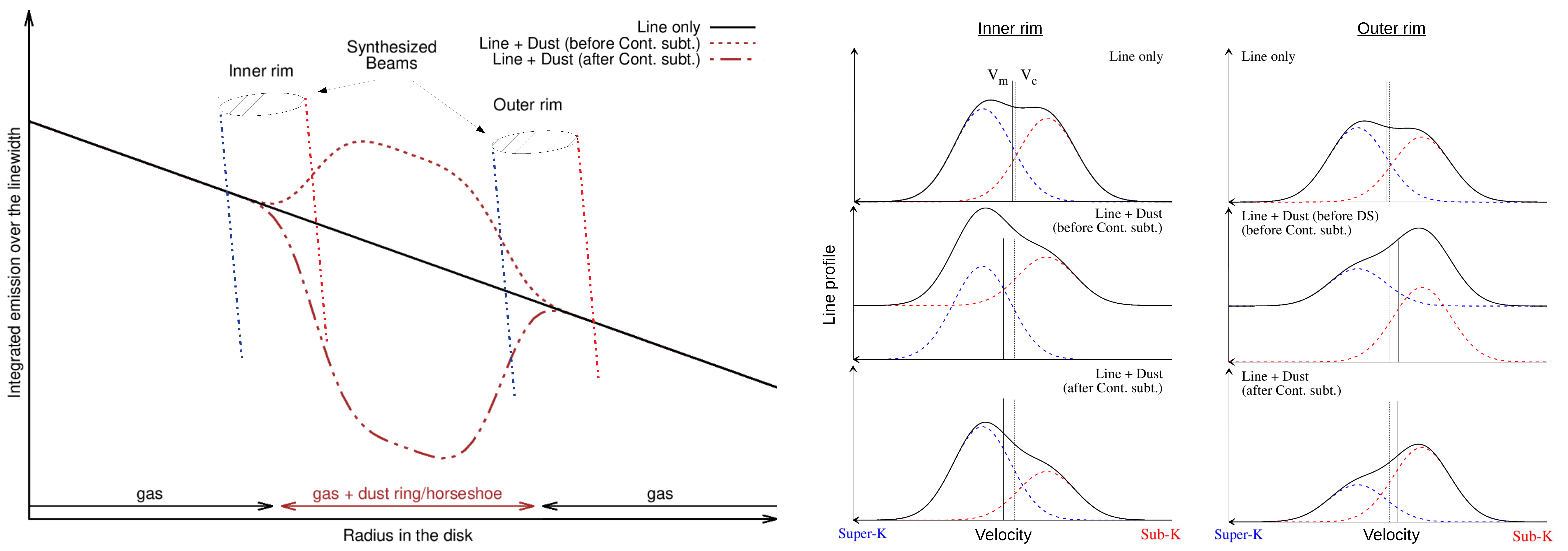}\\
  \caption{Velocity artifacts created by a ring or horseshoe dust structure. Left panel: Sketch representing the integrated gas emission over 
  the line width as a function of the disk radius. The black solid line is the gas emission without the dust ring, and the two red lines 
  represent the emission in presence of the dust ring before and after continuum subtraction. The two ellipses indicate the position of 
  two synthesized beams. Right panel: Line emission profile (black solid line) in the simplified case where the line emission is the sum 
  of the emission at the inner (blue dashed line) and outer (red dashed lines) edges of the synthesized beam. The velocity v$_c$, 
  represented by the dotted black vertical line, is the velocity at the center of the beam while v$_m$, represented by the solid 
  black line, is the intensity weighted velocity.}
  \label{Fig:dust-rings}
\end{figure*}

To illustrate the artifacts in the velocity measurement due to dust concentrations, we present in the left panel of Fig.~\ref{Fig:dust-rings} 
a simple sketch of a disk where the gas surface density is slowly decreasing with radius and which contains a dust ring or horseshoe structure. 
If the gas emission is optically thick, this will create a local and sharp decrease in the gas emission when measured after continuum subtraction. 
Two synthesized beams are represented at the inner and outer rim location of the dust structure where flux variations, and then velocity biases, 
are maximum. The right panel of Fig.~\ref{Fig:dust-rings} shows qualitatively how the spectral profile of the emission line is modified at these 
two positions. Without the presence of the 
dust structure, the measured velocity profile is slightly super-Keplerian due to the slow decrease in the gas emission with radius. 
On the contrary, the steep gradient in the line emission at the rims position artificially creates strong deviations to the Keplerian 
rotation, with a super-Keplerian rotation at the inner rim and a sub-Keplerian rotation at the outer rim. We also note that the line 
profile is only modified by an offset after continuum subtraction, leaving the measured velocity unchanged. 

A similar effect on the measured velocity will happen if the line emission coming from the back side of the disk is absorbed by dust 
particles lying on the midplane. The integrated emission at the ring position, and then the weight given to this location, will also be 
attenuated in the measurement of the gas velocity. This effect can, however, be mitigated if the front molecular layer of the disk, usually 
brighter, can be differentiated spectrally, as shown in $^{12}$CO for AS~209 \citep{Teag2018c}. Such an observation, however, will still be 
affected by the continuum subtraction effect if the molecular line is optically thick.

\subsection{Beam smearing and gas pressure gradients may also explain the vortex-like kinematic feature}

The main difficulty in identifying the origin of the radial Doppler-shifted structure at the north of the disk comes from the possible 
confusion with artifacts in the measurement of the velocity and other kinematic signals. We have shown in the top panel of 
Fig.~\ref{Fig:dev-quant} that beam smearing can produce along the north major axis a pattern similar in morphology to the azimuthal velocity 
deviations that a vortex would produce. However, these artifacts are of lower amplitude in our model, with velocities reaching 90 m s$^{-1}$ 
on the inner side and 20 m s$^{-1}$ on the outer side, while we measure about 150-160 m s$^{-1}$ on both sides in the observations. Besides, 
the kinematic structure due to beam smearing is located at 215 au, at the outer edge of the artificial gas cavity due to continuum subtraction,
while the radial Doppler-shift in the observations is centered at 185 au, at the radius where both dust and gas reach their maximum surface 
density. It is also more extended radially, from 85 to 300 au, compared to 135 to 270 au in the models.

It therefore appears unlikely that the beam smearing effect alone can produce the azimuthal velocity deviations. However, azimuthal 
kinematic deviations are also expected on the inner and outer sides of the circumbinary disk in HD~142527 due to radial gas pressure 
gradients \citep{Teag2018a, Yen2020}. This process may produce super-Keplerian velocities at the inner edge of the horseshoe structure 
and sub-Keplerian velocities outward, as for an anticyclonic vortex. On the south side of the disk, possible deviations 
are up to 30 m s$^{-1}$. With a gas surface density three to five times larger on the northern side of the disk, significant non-Keplerian 
motions may appear and add to the beam smearing effect to possibly produce the large velocity deviations that we observe.

Therefore, the presence of a vortex should be more easily constrained through radial motions to not be mistaken with azimuthal motions 
due to steep gas pressure gradients. At the northern side of the circumbinary disk, we observe a shift toward blueshifted emission 
on the order of 50 m s$^{-1}$ at $\theta$ around $\sim$ 45$^{\circ}$-90$^{\circ}$ and $\sim$ 270$^{\circ}$-315$^{\circ}$, as already 
mentioned in Sect.~\ref{sec:pre-analysis}. This blueshifted emission is visible in Figs.~\ref{Fig:vel-qual}, \ref{Fig:vel-azi}, and 
\ref{Fig:vortex-mod} and may 
correspond to inward motions of the gas at $\theta$ $\sim$ 45$^{\circ}$-90$^{\circ}$ and outward motions at $\theta$ $\sim$ 270$^{\circ}$-315
$^{\circ}$, similar to what would be expected for an anticyclonic vortex. However, as shown in 
Fig.~\ref{Fig:dev-mod}, the beam smearing effect also predicts blueshifted emission at these angles due to a global decrease in the 
emission with radii. It is then currently very difficult to assess if the signal is due to a real vortex or is produced by beam smearing 
associated with gas pressure gradients.

In addition, we note that some uncertainties also exist in the comparison of the measured velocity with the Keplerian rotation. First, 
as shown by \cite{Yen2020} in their appendix, uncertainties in the disk inclination of 2$^{\circ}$ may lead to errors in the measured velocity 
of about 25 m s$^{-1}$ at intermediate angles. In our approach, the disk inclination was fixed but its uncertainty from previous measurements 
is on the order of 1$^{\circ}$ \citep{Fuka2013, Yen2020}. Possible errors due to this process are then limited for our measurements and the 
general pattern predicted by \cite{Yen2020} is not visible in the residuals of the velocity deviations. Actually, most of the uncertainties 
in the measurement of the Keplerian rotation in HD~142527 come from the complexity 
of the kinematic signal that presents large deviations from a pure Keplerian profile, in particular at the north of the disk, as shown in 
Fig.~\ref{Fig:dev-quant}. Our fit was done using the same weight to each radii between 100 and 400 au. If we would have given more weight 
to the inner region of the disk due to the higher local signal-to-noise, the blueshifted emission detected at PA around 
45$^{\circ}$-90$^{\circ}$ and 270$^{\circ}$-315$^{\circ}$ would have been reduced. On the contrary, if velocity deviations due to gradients 
in the gas surface density or due to beam smearing are in fact larger on the inner side of the disk, as our model suggests, the true 
Keplerian rotation would be in reality slightly smaller and the blueshifted emission detected at these angles would then be more 
pronounced.



\section{Discussion}
\label{sec:disc}

\subsection{Observations and theories regarding the possible existence of a large vortex around HD 142527} 
\label{sec:hyp-vortex}

Based on $^{13}$CO and C$^{18}$O J=3-2 data at a spatial resolution of 0.3\arcsec, current observations of the gas kinematics in the 
circumbinary disk around HD~142527 is consistent with the existence of a large vortex, even if complementary observations would be necessary 
to distinguish it from other possibilities. When compared with a model, the center of 
this anticyclonic vortex is located at 185 au from the star, at the estimated radius where both dust and gas surface densities reach 
their maximum \citep{Boeh2017}, with maximum velocity deviations of $\sim$ 150-160 m s$^{-1}$ in the azimuthal direction along the major 
axis, meaning $\sim$ 350 m s$^{-1}$ after deprojection. It is extended radially on each side on about two pressure scale heights, and 
azimuthally over $\sim$ 200$^\circ$, yielding an azimuthal-to-radial aspect ratio $\chi$ of $\sim$ 5.

Using $^{12}$CO, $^{13}$CO, and C$^{18}$O J = 2-1 emission lines, \cite{Garg2020} could estimate the morphology of the disk cavity and 
revealed a steep radial gradient in gas surface density at the inner edge of the circumbinary disk, making it a favorable site for vortex 
formation. Radially wide vortices (i.e., larger than a gas pressure scale height radially) have been predicted in simulations that include 
the displacement of the system barycenter due to the lopsided structure of the disk \citep{Mitt2015, Baru2016}. Such vortices should also 
be azimuthally extended, with ratio between 
4 and 6, to withstand the elliptic instability \citep{Lesu2009}. Recently, \cite{Robe2020} performed numerical simulations in cavity-hosting 
disks with the formation of very extended vortices. For vertical aspect ratios H/R $\gtrsim$ 0.13, with H the pressure scale height in the 
disk, they predict maximum velocity deviations projected along the line-of-sight of 150 m s$^{-1}$ for a disk inclination similar to the one 
in HD~142527. This vertical aspect ratio, while slightly larger than usually considered in protoplanetary disks, may correspond to the vertical 
geometry of the circumbinary disk around HD~142527 as the inner rim is directly illuminated by the Herbig star. A large vortex in HD~142527, 
with a kinematic signal similar to the one observed around the horseshoe structure, is then theoretically possible.

Another interesting result of the simulations performed by \cite{Mitt2015} and \cite{Baru2016} is that dust grains can concentrate at a 
different azimuth than the vortex due to the indirect force exerted by the disk self-gravity. In brief, only small grains with a Stokes number 
$\sim$ 0.01 concentrate in the eye of the vortex due to their strong coupling with the gas while larger grains will concentrate generally 
ahead of it, potentially with a difference of 90$^\circ$ in azimuth, and giving a double peak structure to the continuum emission. At first 
glance, it is tempting to connect this scenario with the double-peaked structure observed in the dust emission in the circumbinary disk around 
HD~142527. Nevertheless, in the observations, the maximum in the dust emission is located at PA = 50$^\circ$, 70$^\circ$ behind the azimuthal 
position of the vortex and of the secondary maximum in dust emission, both located at PA $\sim$ -20$^\circ$, contrary to the predictions in 
\cite{Mitt2015} and in \cite{Baru2016}. 

  It is also possible that the presence of the companion, not present in the previous simulations, plays an important role in the location 
of the large dust grains. \cite{Hamm2019, Hamm2021} found that a large vortex characterized by a flat pressure bump would be sensitive to the 
overlapping of spiral density waves. Dust grains will then not concentrate in a small and centered area but in an elongated and complex 
structure, with a possible off-centered peak, a skewness around this peak, or even a double-peaked structure. While elongated, the double 
peaked structure in the continuum emission may, however, only be due to the shadow cast by the inner disk \citep{Verh2011, Mari2015}. Recent 
modelings of the dust distribution using wavelengths between 0.9 and 3 millimeter, and the temperature given by optically thick molecules, 
suggest that dust grains may in fact reach a density maximum near the major axis \citep{Yen2020}, or even at the position of the secondary 
maximum \citep{Soon2019}, at approximately the vortex position. A solid understanding of the dust concentration process in the HD~142527 
system will require a better mix of long wavelengths and high spatial resolution than the 0.30-0.43\arcsec currently available 
at $\lambda$ $>$ 1 mm.

\subsection{Vortex VS binary}

In an ALMA survey of disks presenting large central cavities, \cite{vdM2020} revealed that dust asymmetries were only present in disks 
with a sufficiently low gas surface density. This suggests that local gas pressure maxima are actually common in such disks but that only dust 
grains with a Stokes number $\gtrsim$ 10$^2$ are efficiently dragged into them. One possibility to create a local gas pressure is through a 
vortex, as already 
discussed in Sect.~\ref{sec:hyp-vortex}. While not mutually exclusive, another theory suggests that an azimuthal gas pressure maximum can be formed 
by the interaction between a binary, with a mass ratio q $\gtrsim$ 0.05 with the main star, and the circumbinary disk. Numerical simulations have 
shown that the massive companion would create a large and eccentric cavity, leading to a gas asymmetry in the gas \citep{Shi2012, Dora2016, 
Ragu2017, Pric2018} rotating at Keplerian velocity and able to trap dust particles \citep{Calc2019, Ragu2020}. 

These studies have been performed using a relatively high 
viscosity $\alpha$ $=$ 5$\times$10$^{-3}$, and therefore did not produce vortices as they require $\alpha$ $\lesssim$ 10$^{-3}$ \citep{Zhu2014}. 
It is, however, plausible that both processes can act together at a smaller viscosity. As shown by \cite{Pric2018}, the HD~142527 system is 
particularly well suited for the binary scenario with the only massive companion (q $\sim$ 0.1-0.15) directly imaged in a large cavity 
\citep{vdM2020}. Other systems like AB Aur \citep{Tang2012}, IRS~48 \citep{vdM2013}, or HD~135344B \citep{Cazz2018} also present 
a large cavity with a horseshoe structure and may correspond to this scenario as well. On the contrary, asymmetries located across ring structures 
in the outer regions of the disks, like in MWC~758, HD~143006 and V1247~Orionis, probably require the presence of a vortex to form \citep{Baru2019}. 

The main uncertainty about the formation and viability of vortices comes from the disk viscosity whose measurements are still scarce. To our 
knowledge, no estimation of the disk viscosity in the circumbinary disk around HD~142527 has been performed. Apart from the notable case of 
DM~Tau where a turbulent viscosity $\alpha$ of $\sim$ 0.08 has been measured 
in the upper layers of the disk using $^{12}$CO emission, only upper limits on the order of a few times 10$^{-3}$ were found using the 
turbulent broadening property of the gas emission lines in HD~163296 \citep{Flah2017}, TW~Hya \citep{Flah2018, Teag2018a}, MWC 480 and 
V4046 Sgr \citep{Flah2020}. Other methods, more indirect because integrating hypotheses on the grains properties, have generally suggested 
$\alpha$ values from a few 10$^{-4}$ to a few 10$^{-3}$ by estimating the dust settling degree in HL~Tau \citep{Pint2016}, or by 
using the radial dust dispersion across rings \citep{Dull2018}. 

Simulations performed using a lower viscosity would help to constrain the complex dust concentration process in the HD~142527 disk.
Further observations of the gas kinematics at a higher spatial resolution, for instance of 0.1\arcsec\ instead of 0.3\arcsec, would allow 
velocity biases due to beam smearing to diminish and a possible vortex signal to be more clearly distinguished from the background rotation. 
With a dust 
absorption coefficient $\beta$ estimated to 1.6 by \cite{Yen2020} at the horseshoe position, it will also be judicious to choose a lower 
frequency, for example with the J=2-1 transitions lines of the CO isotopologues, to significantly reduce the specific beam smearing effect 
around the dust crescent. Indeed, the dust optical depth would decrease by a factor of $\sim$ 2, significantly diminishing the dust 
emission and the depth of the artificial gas cavity.



\subsection{Beam smearing with kinks, spirals, and rings}

  During the analysis of the disk kinematics, we pointed out the beam smearing effect due to variations of the line emission inside of 
the synthesized beam and which may lead to a misinterpretation of kinematic signals. Indeed, local variations of a few tens to a few 
hundreds of m s$^{-1}$ in the gas velocity have been used in the recent years to: i) probe radial pressure gradients and then gaps in 
the gas surface density \citep{Teag2018a, Teag2018c}, ii) observe the kinematic signature of spirals \citep{Teag2019b} and iii) to directly 
trace the presence of planets through the observation of Doppler flips or kinks \citep{Pint2018, Pint2019, Casa2019b, Pint2020}. 

Velocity artifacts due to beam smearing present characteristics that may help in distinguishing them from real kinematic features. 
First, these artifacts should appear in regions where the line emission undergoes a steep spatial variation. This may be produced by a change 
in the gas surface density in the temperature, or processes such as photo-dissociation, freeze-out onto grains, chemical reactions, and 
optical depth effects around dust structures. Second, the artifacts will also increase in regions that present a strong gradient in 
the projected velocity. For a disk dominated by Keplerian motions, this will correspond to regions near the major axis with 
biases decreasing in azimuth with $\sim$ cos($\theta)$, and in regions close to the star with biases decreasing with the radius in $\sim$ 
r$^{-1.5}$. In addition, similarly to the projected Keplerian rotation, biases will increase with the disks inclination, in sin($i$), and 
with the system mass, in M$_\odot^{0.5}$.

In comparison, kinks and Doppler flips are very localized features that do not reveal any preferential azimuthal locations \citep{Pint2018, 
Pint2019, Casa2019b, Pint2020}. Nor do they seem to be linked with a specific point-like cavity or source of emission in the lines 
considered. Therefore, these kinematic signals do not correspond to the expected velocity artifacts due to beam smearing but better agree 
with the presence of a planet, located potentially at any azimuthal angle and perturbing the gas motion in the radial, vertical, or azimuthal 
direction \citep{Pint2019}. 

It is more difficult to directly interpret the gas kinematics around rings and spirals as both structures are generally accompanied by 
variations of the line emission, like around the horseshoe structure in HD~142527. The velocity artifacts created across dust rings have been 
shown in Fig~\ref{Fig:dust-rings} and in Sec~\ref{sec:dust}. The possible decrease in the gas emission at the dust ring position, due to optical 
depth effects, will produce the measurement of a super-Keplerian and of a sub-Keplerian velocity at the inner and outer rims of the dust disk, 
respectively. This is similar to what is theoretically expected if dust grains are trapped in a radial pressure maximum \citep{Kana2015, 
Teag2018a}. True kinematic signatures and artifacts being due to azimuthal motions, there will be both mostly visible around the major axis of 
the disk and their differentiation may require a careful analysis. Inversely, spirals would appear as maximum in the gas 
emission and the beam-smearing effect will artificially produce sub-Keplerian and super-Keplerian velocities on the inner and outer edges of 
these structures. However, compared to ring-like structures, spirals can be more easily distinguished by the variation of their kinematic 
signal as a function of the azimuth as significant non-azimuthal motions around spirals are also expected \citep{Pint2019}.




\section{Conclusion}
\label{sec:concl}

We present a study focusing on the gas kinematics in the lopsided disk surrounding the binary system HD~142527 at a spatial resolution 
of 0.3\arcsec\ and at a wavelength of $\sim$ 0.9 millimeter. Our major findings are:

\begin{enumerate}
    \item The main kinematic structure has a vortex-like morphology with potential azimuthal and radial motions. It is located at the north 
of the circumbinary disk, around the dust concentration, and has projected velocities up to 150 m s$^{-1}$ compared to the Keplerian background.  
Three spiral-like kinematic features are also detected in the outer region of the circumbinary disk at R between 200 and 450 au. Two of them 
are probably related to the two most innermost spirals detected in this system (S1 and S4 following \cite{Garg2020}).    
    \item Using comparisons with a vortex model, the main kinematic feature is consistent with the existence of a large vortex located 
at R $\sim$ 185 au, at the radius where dust and gas reach their maximum surface densities, and at a PA of $\sim$ -20$^{\circ}$. Such a 
vortex may have formed through the Rossby-wave instability at the inner edge of the circumbinary disk, suggesting a relatively low turbulent 
viscosity $\alpha$ $\lesssim$ 10$^{-3}$ \citep{Zhu2014}. Its very large size, of about $\pm$ 40 au radially and $\sim$ 200$^\circ$ azimuthally, 
suggests that the disk self-gravity plays an important role through the indirect force \citep{Mitt2015, Baru2016}. The relatively large 
velocities of approximately 350 m s$^{-1}$ after deprojection, comparable to the local sound speed, have been predicted in simulations of 
vortices performed by \cite{HuanP2018} and \cite{Robe2020}.
    \item Velocity measurements are, however, subject to artifacts due to variations of the line emission at a sub-synthesized beam scale. 
If associated with non-Keplerian motions due to gas pressure gradients at the inner and outer edges of the circumbinary disk, the beam smearing 
effect may also create a kinematic signal similar to an anticyclonic vortex around the horseshoe structure. Our current observations, at a 
spatial resolution of 0.3\arcsec, do not allow us to distinguish between these two possibilities.
    \item Velocity deviations due to beam smearing should be common in protoplanetary disks and may lead to the misinterpretations of kinematic 
signals, in particular around ring-like structures. The principal method to reduce them is to perform observations at a higher spatial resolution. 
This would also allow, through a better knowledge of the disk structure, to constrain much more precisely the amplitude and localization of such 
artifacts. In HD~142527, a spatial resolution of 0.1\arcsec\ is reachable in 2-3 hours of telescope time on target and would decrease by a 
factor of $\sim$ 3 the current beam smearing effect along the major axis of the disk. 
\end{enumerate}

\begin{acknowledgements} 
We thank the referee whose comments improved the quality of the manuscript, and G. Lesur for useful discussions about vortex properties. The 
authors acknowledge funding from ANR 
(Agence Nationale de la Recherche) of France under contract number ANR-16-CE31-0013 (Planet-Forming-Disks). JFG thanks the LABEX 
Lyon Institute of Origins (ANR-10-LABX-0066) of the Universit\'e de Lyon for its financial support within the programme 
`Investissements d'Avenir' (ANR-11-IDEX-0007) of the French government operated by the ANR. This paper makes use of the following 
ALMA data: ADS/JAO.ALMA\#2012.1.00725.S. ALMA is a partnership of ESO (representing its member states), NSF (USA) and NINS (Japan), 
together with NRC (Canada), MOST and ASIAA (Taiwan), and KASI (Republic of Korea), in cooperation with the Republic of Chile. The 
Joint ALMA Observatory is operated by ESO, AUI/NRAO and NAOJ.

This work has made use of data from the European Space Agency (ESA) mission {\it Gaia} (\url{https://www.cosmos.esa.int/gaia}), processed 
by the {\it Gaia} Data Processing and Analysis Consortium (DPAC,\url{https://www.cosmos.esa.int/web/gaia/dpac/consortium}). Funding for 
the DPAC has been provided by national institutions, in particular the institutions participating in the {\it Gaia} Multilateral 
Agreement.
\end{acknowledgements}

\begin{appendix}
\section{A 3D toy model for HD~142527} 
\label{app:model}

\begin{figure*}[!h]
  \includegraphics[angle=0,width=\textwidth]{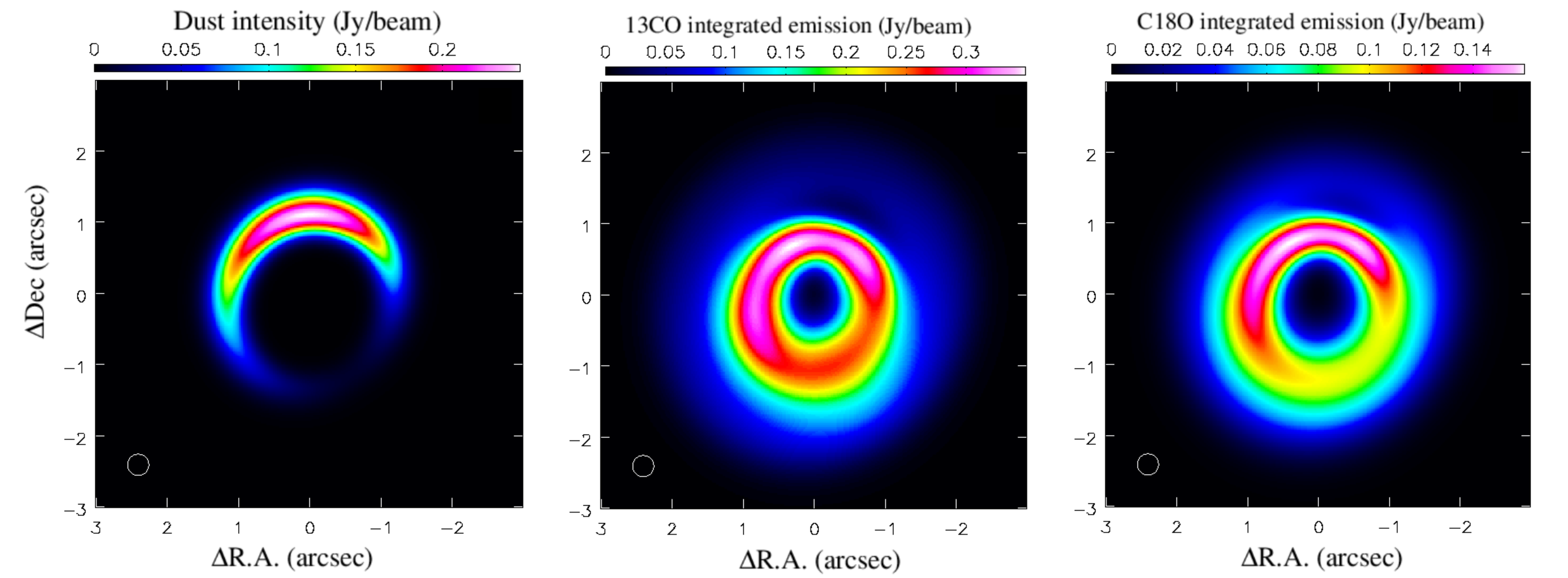}\\
  \caption{From the left to the right: Dust emission, $^{13}$CO J = 3-2 and C$^{18}$O J = 3-2 integrated emission of the 3D toy model of HD~142527. The spatial resolution is 0.3\arcsec.}
  \label{Fig:model-image}
\end{figure*}

The model of the circumbinary disk around HD~142527 is built in a spherical framework ($r$, $\theta$, $\phi$) using cells of 1 au in radius, 
2$^\circ$ in azimuth and 1$^\circ$ in elevation. The horseshoe structure in the dust and gas surface density is 
modeled using modified Gaussians along the radial and azimuthal directions following the formula:
\textbf{
\begin{equation}
  \begin{aligned}                                  
  \Sigma(r, \theta) = {} & (\Sigma_\mathrm{max}-\Sigma_\mathrm{min})~\mathrm{exp}\left[ -\left( \frac{r-R_0(\theta)}{w_\mathrm{in,out}(\theta)} 
  \right)^2 \right]~\mathrm{exp}\left[ -\left( \frac{\theta-\theta_\mathrm{max}}{w_\mathrm{ctcl,cl}} \right)^2 \right]  \\  
                         & + \Sigma_\mathrm{min}, 
  \end{aligned} 
  \label{eq:mod-rad}
\end{equation}}
with $\Sigma_{max}$ the surface density at the position ($R_\mathrm{0}(\theta_\mathrm{max})$, $\theta_\mathrm{max}$), where the dust and 
gas surface densities are maximum, and $\Sigma_{min}$ the surface density at the position ($R_\mathrm{0}(\theta_\mathrm{min})$, 
$\theta_\mathrm{min}$) for the azimuth where the gas and dust surface densities are minimum. The parameters $w_\mathrm{in}(\theta)$ and 
$w_\mathrm{out}(\theta)$ 
are the radial half-widths of the Gaussian for $r$ $<$ $R_\mathrm{0}$ or for $r$ $>$ $R_\mathrm{0}$, and $w_\mathrm{ctcl}$ and 
$w_\mathrm{cl}$ the azimuthal half-widths of the Gaussian in the counterclockwise and clockwise directions, starting from $\theta_\mathrm{max}$. 

To represent the eccentric aspect of the circumbinary disk, the parameters $R_\mathrm{0}(\theta)$, and $w_\mathrm{in,out}(\theta)$ vary as a 
function of the azimuth through the formula, for instance for $R_\mathrm{0}(\theta)$ :
\begin{equation}
   \begin{aligned}
   R_0(\theta)  = & ~ ( R_0(\theta_\mathrm{max})- R_0(\theta_\mathrm{min})) ~ \mathrm{exp}\left[ -\left( \frac{\theta-\theta_\mathrm{max}}
   {w_{ctcl, cl}} \right)^2 \right] + R_0(\theta_\mathrm{min}).
  \label{eq:mod-azi}
  \end{aligned}
\end{equation}
The value of the parameters are given in Table~\ref{tab:dens}. They were inspired by the analysis in \cite{Boeh2017}, using the same 
grain properties, but also by the studies performed by \cite{Soon2019} and \cite{Yen2020}, which took into account the azimuthal variation of 
the temperature, obtained using optically thick molecules, and then estimated that the gas surface density maximum was at $\theta$ $\sim$ 
15$^\circ$.  
\begin{table}
\centering
  \begin{tabular}{|ccccc|}
    \hline
         & $\Sigma_{0}$ (g cm$^{-2}$) & $R_0$ (au) &  $w_\mathrm{in}$ (au) & $w_\mathrm{out}$ (au) \\ \hline          
    \multicolumn{5}{l}{Values at $\theta_\mathrm{max}$ = 15$^\circ$} \\ \hline
    Dust & 0.7    & 185  & 26  & 47   \\
    Gas  & 1.2    & 185  & 60  & 85   \\ \hline
    \multicolumn{5}{l}{Values at $\theta_\mathrm{min}$ = 225$^\circ$} \\ \hline
    Dust & 0.012   & 205  & 17  & 43  \\
    Gas  & 0.24   & 205  & 90  & 100  \\ \hline
    \multicolumn{5}{l}{ } \\
    \multicolumn{5}{l}{Azimuthal variation} \\ \hline   
         & \multicolumn{2}{c}{$w_\mathrm{ctcl}$ (au)} & \multicolumn{2}{c}{$w_\mathrm{cl}$ (au)} \\ \hline
    Dust \& Gas & \multicolumn{2}{c}{95}  & \multicolumn{2}{c}{65}  \\ \hline       
  \end{tabular}
  \caption{Density prescription for the Gaussians describing the dust and gas surface density on the northern and southern profiles.}     
  \label{tab:dens}
\end{table}
We then calculated the circumbinary disk temperature through radiative transfer by using RADMC-3D \citep{Dull2012}. No shadowing effect by the 
inner disk has been taken into account in this step. With the same code, we performed the ray-tracing in a square grid with a pixel size of 
20 mas and a velocity resolution of 10 m s$^{-1}$, about ten times smaller than in our observations, to produce the final image before 
beam dilution. In this model, the gas is in pure Keplerian rotation around a single star of 2.36 $M_\odot$ with $v(r,z)$ $=$ 
$\sqrt{GM/(r^2 + z^2)^{0.5}}$. We assume a disk inclination of 27$^\circ$ and a PA of -19$^\circ$. Images of the circumbinary disk 
model for the dust intensity, and the $^{13}$CO and C$^{18}$O J=3-2 integrated emission are given in Fig.~\ref{Fig:model-image}.


\section{Comparison of the ``peak'' and ``intensity weighted'' methods}
\label{sec:mom9}

We have used in this study the intensity weighted method. Another popular method to estimate the gas velocity along the line-of-sight 
is the peak emission method. It consists in identifying for each pixel of the map the spectral position of the peak of the line emission. 
The precision of this method has recently been improved in \cite{Teag2018b} by performing a polynomial fit on the three channels around 
the maximum in emission. This method is especially useful when the front and back molecular layers can be disentangled spatially and 
spectrally (see also Appendix A.3 in \cite{Teag2018c}). This allows the precise determination of the origin ($r$, $\theta$, $z$) of the emission 
and the comparison of the front layer of the disk with the Keplerian velocity. However, such observations are still rare. In our observations, 
the low inclination of the circumbinary disk around HD~142527, the choice of the $^{13}$CO and C$^{18}$O transitions lines that emit 
at a lower altitude than $^{12}$CO, and the moderate spatial resolution of our observations, do not give us the ability to distinguish the 
two molecular layers.  

We mainly privileged the intensity weighted method in our study because it appeared less affected by the rms noise than the peak 
emission approach, in particular with the $^{13}$CO J=3-2 transition line. This is probably due to the high optical depth of the 
line and to the moderate spatial resolution of our observations, which give to the peak of the emission a flattened aspect. Another 
reason for the better precision of the intensity weighted method within our data is the good signal-to-noise in the circumbinary 
disk where about 10 channels are involved in the measurement of the gas velocity.

\begin{figure*}[ht]
  \includegraphics[angle=-90,width=\textwidth]{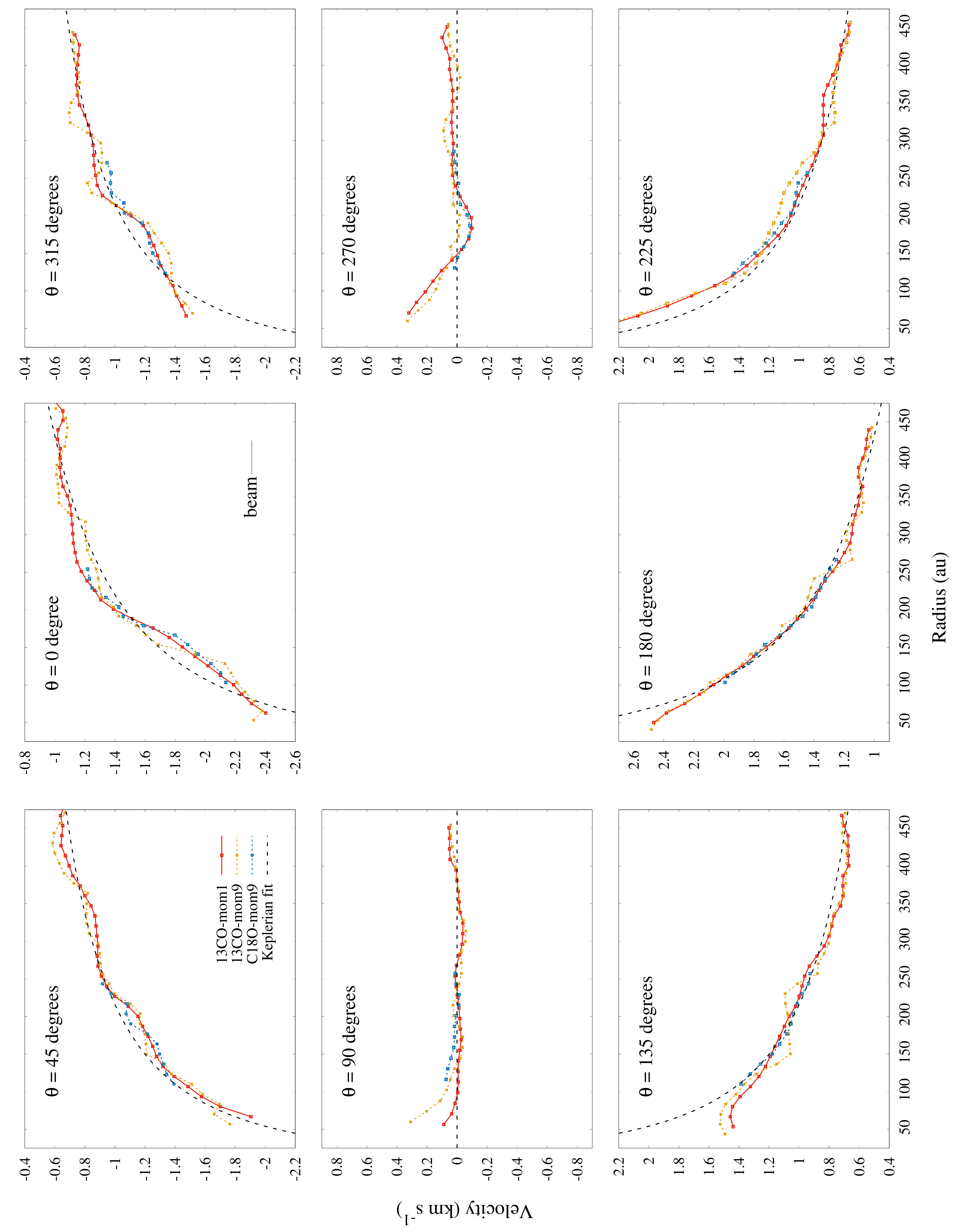}\\
  \caption{Radial profiles of the velocity for different PAs $\theta$, spaced by 45$^{\circ}$, where $\theta$ is the angle 
  starting from the northern major axis and rotating counterclockwise. The red curve is the velocity measured using the intensity weighted 
  method (mom1 in CASA) with $^{13}$CO J=3-2. The orange and blue curves correspond to the velocity measured using the peak emission method 
  with the $^{13}$CO J=3-2 and C$^{18}$O J=3-2 transition lines, respectively, with the procedure detailed in \cite{Teag2018b}.}
  \label{Fig:vel-mom9}
\end{figure*}

We show in Fig.~\ref{Fig:vel-mom9} the velocity measured from the $^{13}$CO and C$^{18}$O J=3-2 transition lines using the intensity 
weighted and the peak emission methods. For the intensity weighted method, we only indicate the velocity measured in 
$^{13}$CO as both molecules show the same velocity (cf Fig.~\ref{Fig:vel-azi}). The results between both methods and molecules are 
in general agreement, with a Keplerian profile at the south of the disk and a super-Keplerian profile at the north for R $\lesssim$ 
160-180 au and sub-Keplerian for R $\gtrsim$ 160-180 au. 

There are also slight differences between the methods and the lines involved, which is in itself not a surprising result. 
First, the two methods would only give the same results if the line profile was symmetric around the peak of emission, what may not 
be the case in practice. For instance, spectral profiles similar to the examples shown in the right panel of Fig.~\ref{Fig:dust-rings} 
would give very different values. Second, the two methods do not probe the same regions ($r$, $\theta$, $z$) along the line-of-sight 
due to optical depth effects. In fact, the largest differences in the measured velocity in Fig.~\ref{Fig:vel-mom9} are found between 
the peak emission of the $^{13}$CO line, optically thick and only probing the front $^{13}$CO layer, and the two other measurements, 
which are the peak intensity method applied to the moderately optically thick C$^{18}$O and the intensity weighted method, which probes 
all the line profile including optically thin wings, therefore also sensitive to the back molecular layer. Finally, the optical 
depth can also modify the surface brightness, and thus the velocity artifacts due to beam smearing, as the weight given to each region 
of the synthesized beam would change. 

\end{appendix}


\begin{thebibliography}{99}  
\bibitem[Andrews et al.(2018)]{Andr2018} Andrews, S.~M., Huang, J., P{\'e}rez, L.~M., et al.\ 2018, \apjl, 869, L41 
\bibitem[Auffinger \& Laibe(2018)]{Auff2018} Auffinger, J., \& Laibe, G.\ 2018, \mnras, 473, 796 
\bibitem[Bai \& Stone(2010)]{Bai2010} Bai, X.-N., \& Stone, J.~M.\ 2010, \apj, 722, 1437 
\bibitem[Barge \& Sommeria(1995)]{Barg1995} Barge, P., \& Sommeria, J.\ 1995, \aap, 295, L1 
\bibitem[Baruteau \& Zhu(2016)]{Baru2016} Baruteau, C., \& Zhu, Z.\ 2016, \mnras, 458, 3927 
\bibitem[Baruteau et al.(2019)]{Baru2019} Baruteau, C., Barraza, M., P{\'e}rez, S., et al.\ 2019, \mnras, 486, 304
\bibitem[Biller et al.(2012)]{Bill2012} Biller, B., Lacour, S., Juh{\'a}sz, A., et al.\ 2012, \apjl, 753, L38
\bibitem[Birnstiel et al.(2010)]{Birn2010} Birnstiel, T., Dullemond, C.~P., \& Brauer, F.\ 2010, \aap, 513, A79 
\bibitem[Boehler et al.(2017)]{Boeh2017} Boehler, Y., Weaver, E., Isella, A., et al.\ 2017, \apj, 840, 60
\bibitem[Boehler et al.(2018)]{Boeh2018} Boehler, Y., Ricci, L., Weaver, E., et al.\ 2018, \apj, 853, 162
\bibitem[Calcino et al.(2019)]{Calc2019} Calcino, J., Price, D.~J., Pinte, C., et al.\ 2019, \mnras, 490, 2579 
\bibitem[Casassus et al.(2013)]{Casa2013} Casassus, S., van der Plas, G., M, S.~P., et al.\ 2013, \nat, 493, 191 
\bibitem[Casassus et al.(2015)]{Casa2015} Casassus, S., Marino, S., P{\'e}rez, S., et al.\ 2015, \apj, 811, 92 
\bibitem[Casassus et al.(2019)]{Casa2019a} Casassus, S., Marino, S., Lyra, W., et al.\ 2019, \mnras, 483, 3278 
\bibitem[Casassus \& P{\'e}rez(2019)]{Casa2019b} Casassus, S., \& P{\'e}rez, S.\ 2019, \apjl, 883, L41 
\bibitem[Cazzoletti et al.(2018)]{Cazz2018} Cazzoletti, P., van Dishoeck, E.~F., Pinilla, P., et al.\ 2018, \aap, 619, A161 
\bibitem[Chavanis(2000)]{Chav2000} Chavanis, P.~H.\ 2000, \aap, 356, 1089
\bibitem[Christiaens et al.(2014)]{Chri2014} Christiaens, V., Casassus, S., Perez, S., et al.\ 2014, \apjl, 785, L12
\bibitem[Claudi et al.(2019)]{Clau2019} Claudi, R., Maire, A.-L., Mesa, D., et al.\ 2019, \aap, 622, A96 
\bibitem[de Val-Borro et al.(2007)]{deVa2007} de Val-Borro, M., Artymowicz, P., D'Angelo, G., et al.\ 2007, \aap, 471, 1043
\bibitem[Dong et al.(2018)]{Dong2018} Dong, R., Liu, S.-y., Eisner, J., et al.\ 2018, \apj, 860, 124 
\bibitem[D'Orazio et al.(2016)]{Dora2016} D'Orazio, D.~J., Haiman, Z., Duffell, P., MacFadyen, A., \& Farris, B.\ 2016, \mnras, 459, 2379 
\bibitem[Dullemond et al.(2012)]{Dull2012} Dullemond, C.~P., Juhasz, A., Pohl, A., et al.\ 2012, Astrophysics Source Code Library, ascl:1202.015 
\bibitem[Dullemond et al.(2018)]{Dull2018} Dullemond, C.~P., Birnstiel, T., Huang, J., et al.\ 2018, \apjl, 869, L46 
\bibitem[Flaherty et al.(2017)]{Flah2017} Flaherty, K.~M., Hughes, A.~M., Rose, S.~C., et al.\ 2017, \apj, 843, 150. 
\bibitem[Flaherty et al.(2018)]{Flah2018} Flaherty, K.~M., Hughes, A.~M., Teague, R., et al.\ 2018, \apj, 856, 117 
\bibitem[Flaherty et al.(2020)]{Flah2020} Flaherty, K., Hughes, A.~M., Simon, J.~B., et al.\ 2020, \apj, 895, 109 
\bibitem[Fukagawa et al.(2013)]{Fuka2013} Fukagawa, M., Tsukagoshi, T., Momose, M., et al.\ 2013, \pasj, 65, L14
\bibitem[Gaia Collaboration et al.(2018)]{Gaia2018} Gaia Collaboration, Brown, A.~G.~A., Vallenari, A., et al.\ 2018, \aap, 616, A1
\bibitem[Garg et al.(2020)]{Garg2020} Garg, H., Pinte, C., Christiaens, V., et al.\ 2020, arXiv:2010.15310
\bibitem[Goodman et al.(1987)]{Good1987} Goodman, J., Narayan, R., \& Goldreich, P.\ 1987, \mnras, 225, 695
\bibitem[Hammer et al.(2019)]{Hamm2019} Hammer, M., Pinilla, P., Kratter, K.~M., et al.\ 2019, \mnras, 482, 3609
\bibitem[Hammer et al.(2021)]{Hamm2021} Hammer, M., Lin, M.-K., Kratter, K.~M., et al.\ 2021, arXiv:2104.02782
\bibitem[Huang et al.(2018)]{HuanP2018} Huang, P., Isella, A., Li, H., et al.\ 2018, \apj, 867, 3. 
\bibitem[Isella et al.(2018)]{Isel2018} Isella, A., Huang, J., Andrews, S.~M., et al.\ 2018, \apjl, 869, L49
\bibitem[Johansen \& Youdin(2007)]{Joha2007} Johansen, A., \& Youdin, A.\ 2007, \apj, 662, 627 
\bibitem[Kanagawa et al.(2015)]{Kana2015} Kanagawa, K.~D., Tanaka, H., Muto, T., et al.\ 2015, \mnras, 448, 994
\bibitem[Keppler et al.(2019)]{Kepp2019} Keppler, M., Teague, R., Bae, J., et al.\ 2019, \aap, 625, A118
\bibitem[Kida(1981)]{Kida1981} Kida, S.\ 1981, Journal of the Physical Society of Japan, 50, 3517 
\bibitem[Kraus et al.(2017)]{Krau2017} Kraus, S., Kreplin, A., Fukugawa, M., et al.\ 2017, \apjl, 848, L11. 
\bibitem[Lesur \& Papaloizou(2009)]{Lesu2009} Lesur, G. \& Papaloizou, J.~C.~B.\ 2009, \aap, 498, 1.  
\bibitem[Li et al.(2000)]{Li2000} Li, H., Finn, J.~M., Lovelace, R.~V.~E., et al.\ 2000, \apj, 533, 1023 
\bibitem[Lin(2012)]{Lin2012} Lin, M.-K.\ 2012, \mnras, 426, 3211
\bibitem[Liu et al.(2018)]{Liu2018} Liu, S.-F., Jin, S., Li, S., et al.\ 2018, \apj, 857, 87. 
\bibitem[Lovelace et al.(1999)]{Love1999} Lovelace, R.~V.~E., Li, H., Colgate, S.~A., \& Nelson, A.~F.\ 1999, \apj, 513, 805 
\bibitem[Lyra \& Lin(2013)]{Lyra2013} Lyra, W., \& Lin, M.-K.\ 2013, \apj, 775, 17 
\bibitem[Marino et al.(2015)]{Mari2015} Marino, S., Perez, S., \& Casassus, S.\ 2015a, \apjl, 798, L44 
\bibitem[McMullin et al.(2007)]{McMu2007} McMullin, J.~P., Waters, B., Schiebel, D., et al.\ 2007, Astronomical Data Analysis Software 
and Systems XVI, 376, 127
\bibitem[Mittal \& Chiang(2015)]{Mitt2015} Mittal, T., \& Chiang, E.\ 2015, \apjl, 798, L25
\bibitem[Muto et al.(2015)]{Muto2015} Muto, T., Tsukagoshi, T., Momose, M., et al.\ 2015, \pasj, 67, 122 
\bibitem[Pinte et al.(2016)]{Pint2016} Pinte, C., Dent, W.~R.~F., M{\'e}nard, F., et al.\ 2016, \apj, 816, 25. 
\bibitem[Pinte et al.(2018)]{Pint2018} Pinte, C., Price, D.~J., M{\'e}nard, F., et al.\ 2018, \apjl, 860, L13 
\bibitem[Pinte et al.(2019)]{Pint2019} Pinte, C., van der Plas, G., M{\'e}nard, F., et al.\ 2019, Nature Astronomy, 3, 1109 
\bibitem[Pinte et al.(2020)]{Pint2020} Pinte, C., Price, D.~J., M{\'e}nard, F., et al.\ 2020, \apjl, 890, L9 
\bibitem[Price et al.(2018)]{Pric2018} Price, D.~J., Cuello, N., Pinte, C., et al.\ 2018, \mnras, 477, 1270 
\bibitem[Rab et al.(2020)]{Rab2020} Rab, C., Kamp, I., Dominik, C., et al.\ 2020, \aap, 642, A165
\bibitem[Raettig et al.(2015)]{Raet2015} Raettig, N., Klahr, H., \& Lyra, W.\ 2015, \apj, 804, 35 
\bibitem[Ragusa et al.(2017)]{Ragu2017} Ragusa, E., Dipierro, G., Lodato, G., Laibe, G., \& Price, D.~J.\ 2017, \mnras, 464, 1449 
\bibitem[Ragusa et al.(2020)]{Ragu2020} Ragusa, E., Alexander, R., Calcino, J., et al.\ 2020, \mnras, 499, 3362
\bibitem[Reg{\'a}ly et al.(2012)]{Rega2012} Reg{\'a}ly, Z., Juh{\'a}sz, A., S{\'a}ndor, Z., \& Dullemond, C.~P.\ 2012, \mnras, 419, 1701 
\bibitem[Richard et al.(2013)]{Rich2013} Richard, S., Barge, P., \& Le Diz{\`e}s, S.\ 2013, \aap, 559, A30. 
\bibitem[Robert et al.(2020)]{Robe2020} Robert, C.~M.~T., M{\'e}heut, H., \& M{\'e}nard, F.\ 2020, \aap, 641, A128.
\bibitem[Rosotti et al.(2020)]{Roso2020} Rosotti, G.~P., Teague, R., Dullemond, C., et al.\ 2020, \mnras, 495, 173. 
\bibitem[Shakura \& Sunyaev(1973)]{Shak1973} Shakura, N.~I. \& Sunyaev, R.~A.\ 1973, \aap, 500, 33
\bibitem[Shi et al.(2012)]{Shi2012} Shi, J.-M., Krolik, J.~H., Lubow, S.~H., \& Hawley, J.~F.\ 2012, \apj, 749, 118 
\bibitem[Sierra et al.(2017)]{Sier2017} Sierra, A., Lizano, S., \& Barge, P.\ 2017, \apj, 850, 115
\bibitem[Soon et al.(2019)]{Soon2019} Soon, K.-L., Momose, M., Muto, T., et al.\ 2019, \pasj, 71, 124 
\bibitem[Surville \& Barge(2015)]{Surv2015} Surville, C., \& Barge, P.\ 2015, \aap, 579, A100 
\bibitem[Tang et al.(2012)]{Tang2012} Tang, Y.-W., Guilloteau, S., Pi{\'e}tu, V., et al.\ 2012, \aap, 547, A84 
\bibitem[Teague et al.(2018a)]{Teag2018a} Teague, R., Bae, J., Bergin, E.~A., Birnstiel, T., \& Foreman-Mackey, D.\ 2018, \apjl, 860, L12 
\bibitem[Teague \& Foreman-Mackey(2018b)]{Teag2018b} Teague, R., \& Foreman-Mackey, D.\ 2018, Research Notes of the American Astronomical Society, 2, 173 
\bibitem[Teague et al.(2018c)]{Teag2018c} Teague, R., Bae, J., Birnstiel, T., \& Bergin, E.~A.\ 2018, \apj, 868, 113 
\bibitem[Teague(2019)]{Teag2019a} Teague, R.\ 2019, The Journal of Open Source Software, 4, 1220. 
\bibitem[Teague et al.(2019b)]{Teag2019b} Teague, R., Bae, J., Huang, J., et al.\ 2019, \apjl, 884, L56
\bibitem[van der Marel et al.(2013)]{vdM2013} van der Marel, N., van Dishoeck, E.~F., Bruderer, S., et al.\ 2013, Science, 340, 1199 
\bibitem[van der Marel et al.(2015)]{vdM2015} van der Marel, N., Pinilla, P., Tobin, J., et al.\ 2015, \apjl, 810, L7 
\bibitem[van der Marel et al.(2016)]{vdM2016} van der Marel, N., Cazzoletti, P., Pinilla, P., et al.\ 2016, \apj, 832, 178 
\bibitem[van der Marel et al.(2018)]{vdM2018} van der Marel, N., Williams, J.~P., Ansdell, M., et al.\ 2018, \apj, 854, 177. 
\bibitem[van der Marel et al.(2020)]{vdM2020} van der Marel, N., Birnstiel, T., Garufi, A., et al.\ 2020, arXiv:2010.10568 
\bibitem[Varni{\`e}re \& Tagger(2006)]{Varn2006} Varni{\`e}re, P., \& Tagger, M.\ 2006, \aap, 446, L13  
\bibitem[Verhoeff et al.(2011)]{Verh2011} Verhoeff, A.~P., Min, M., Pantin, E., et al.\ 2011, \aap, 528, A91 
\bibitem[Weaver et al.(2018)]{Weav2018} Weaver, E., Isella, A., \& Boehler, Y.\ 2018, \apj, 853, 113
\bibitem[Weidenschilling(1977)]{Weid1977} Weidenschilling, S.~J.\ 1977, \mnras, 180, 57 
\bibitem[Yen \& Gu(2020)]{Yen2020} Yen, H.-W. \& Gu, P.-G.\ 2020, arXiv:2010.13990 
\bibitem[Youdin \& Goodman(2005)]{Youd2005} Youdin, A.~N., \& Goodman, J.\ 2005, \apj, 620, 459 
\bibitem[Zhu \& Stone(2014b)]{Zhu2014} Zhu, Z., \& Stone, J.~M.\ 2014, \apj, 795, 53  

\end{thebibliography}
\end{document}